# Pre-accretional irradiation of comet 67P/Churyumov-Gerasimenko constituents

Inès Sansberro[1], Nicolas Fray[1], Hervé Cottin[1], Cécile Engrand[2], Donia Baklouti[3], Christelle Briois[4], Anaïs Bardyn[1,4], Henning Fischer[5], Klaus Hornung[6], John Paquette[7,8], Jouni Ryno[9], Olivier J. Stenzel[5], Laurent Thirkell[4] and Martin Hilchenbach[5*]

(1) Univ Paris Est Creteil and Université Paris Cité, CNRS, LISA, F-94010 Créteil, France,
(2) IJCLab, CNRS, Université Paris-Saclay, 91405 Orsay Cedex, France,
(3) Institut d'Astrophysique Spatiale, CNRS, Université Paris-Saclay, 91405 Orsay Cedex, France,
(4) LPC2E, OSUC, Univ Orleans, CNRS, CNES, F-45071 Orleans, France
(5) Max-Planck-Institut für Sonnensystemforschung, Justus-von-Liebig-Weg 3, 37077, Göttingen, Germany,
(6) Universität der Bundeswehr München, LRT-7, 85577, Neubiberg, Germany
(7) NASA-Goddard Space Flight Center Solar System Exploration Division, Code 690 Greenbelt, MD 20771
(8) Catholic University of America, 620 Michigan Ave., Washington, DC 20064, USA
(9) Finnish Meteorological Institute, Erik Palmenin aukio 1, P.O.Box 503, 00101, Helsinki, Finland

*Deceased

# Abstract

Comets are thought to have largely preserved the primordial material from which they formed. Therefore, cometary studies provide important clues about the processes involved in the formation of the Solar System. During the last weeks of ESA's Rosetta mission, the COSIMA (COmetary Secondary Ion Mass Analyser) instrument performed a series of measurements, which involved long-term sputtering of dust particles captured from comet 67P/Churyumov-Gerasimenko. These post-sputtering analyses revealed the subsurface composition of 8 particles, which we compare here with surfaces composition measured on 59 un-sputtered particle fragments for 9 elements. We found that the analysed surfaces are strongly depleted in metals, such as Mg and Fe, compared to solar abundances, while the subsurface material is consistent with solar values. Because the dust captured by COSIMA fragmented upon impact with the collection targets, this metal depletion cannot reside exclusively on the outer surface of the bulk particles. Instead, these depletion layers must coat the surface of smaller, unfragmented mineral subunits. Such depletion layers, which are often called 'rims', are common at the surface of small regolith dust particles found on airless bodies in our solar system and are attributed to space weathering. We suggest that these mineral cometary subunits experienced pre-accretional solar wind irradiation in the inner regions of the solar nebula during the early stages of Solar System formation. The small irradiated dust constituents must then have been transported to the outer regions of the solar nebula, where they mixed with ices to form the nucleus of the comet.

# 1 INTRODUCTION

Comets are small bodies of the Solar System. Due to their small sizes and their formation in a cold environment, far from the Sun, the composition of comets provides key constraints on the physico-chemical processes prevailing in the protoplanetary disc and its composition (Willacy *et al.* 2015). They are therefore expected to contain more light elements than the most primitive objects in the asteroid belt, which are the parent bodies of carbonaceous chondrite meteorites (Bell *et al.* 1989). The dust particles ejected from cometary nuclei should therefore be among the most primitive materials in the Solar System, having remained virtually unaltered and preserved at low temperatures since their co-accretion with ices to form the cometary icy bodies (Mumma & Charnley 2011).

Several past space missions have provided initial insights into the composition of cometary dust. In 1986, the first in-situ analyses of dust particles from Comet 1P/Halley were carried out as part of the Giotto, Vega 1 and Vega 2 missions. On board these three missions, mass spectrometers were devoted to measuring the elemental composition of cometary particles ejected from the nucleus. The large relative velocities between 1P/Halley and these probes (68 km $s^{-1}$, 79 km $s^{-1}$ and 77 km $s^{-1}$ for Giotto, Vega 1 and Vega 2, respectively) allowed for direct ionisation by impact, leading to the measurement of the bulk elemental composition of 1P/Halley's dust particles. These missions revealed that cometary dust particles contain minerals as well as carbonaceous materials composed of C, H, O and N atoms (Kissel *et al.* 1986a, Kissel *et al.* 1986b, Clark *et al.* 1987, Jessberger *et al.* 1988, Lawler & Brownlee 1992, Jessberger 1999). Their overall composition was estimated to be similar to that of the solar photosphere for most of the elements measured. In particular, the overall C/Si atomic ratio of particles from 1P/Halley was estimated to be 4.4 ± 1.3, nearly five times the average C/Si ratio of 0.94 ± 0.18 observed in CI-type carbonaceous chondrite meteorites, and less than half the protosolar abundance of 9.12 ± 0.18 (Lodders 2021). Twenty years later, the Stardust mission enabled the first laboratory analysis of samples brought back to Earth from 81P/Wild 2, a Jupiter family comet. One of the main conclusions was that the returned particles contained crystalline minerals (mainly silicates) that had been processed at high temperatures in the inner regions of the protoplanetary disc (Brownlee 2014). However, quantifying the native carbonaceous matter from 81P/Wild 2 was challenging due to the high-velocity impact (6.1 km $s^{-1}$) into silica aerogel that already contained background carbon contamination, and the resulting thermal alteration of the captured grains during impact (Brownlee 2014, Sandford *et al.* 2010).

In addition to in-situ space missions, thermal infrared spectroscopy allows for the study of cometary particle composition for a larger number of comets (Harker *et al.* 2023, Kolokolova *et al.* 2004, Kolokolova *et al.* 2024). These remote observations confirm that cometary dust contains, on average, about 50% carbonaceous material by mass and show that the minerals are both amorphous and crystalline. Moreover, detailed studies of the 10 µm and 18 µm thermal infrared spectral features suggest prior ion irradiation of amorphous Mg-Fe silicates before their incorporation into comets (Harker et al. 2023).

The similarities of the cometary particles to the anhydrous chondritic aggregate class of interplanetary dust particles (IDPs) suggest that comets may be the source of these IDPs (Engrand *et al.* 2024, Hanner & Bradley 2004, Flynn *et al.* 2016). In particular, these porous

IDPs, which offer the advantage of being studied with high-sensitivity laboratory tools, exhibit higher carbon abundances than CI chondrites (Thomas *et al.* 1993, Engrand *et al*. 2024).

Throughout the Rosetta mission of the European Space Agency (ESA), from August 2014 to September 2016, the mass spectrometer COSIMA (COmetary Secondary Ion Mass Analyzer) collected, at low relative velocities (a few m $s^{-1}$) (Della Corte *et al.* 2015), about 35,000 dust particle fragments ejected from the nucleus of comet 67P/Churyumov-Gerasimenko (hereafter 67P) (Merouane *et al.* 2017). Once collected on the sampling targets, dust particle fragments could be imaged, and about 250 of them were analysed by COSIMA (Langevin *et al.* 2016, Bardyn *et al.* 2017). COSIMA is a time-of-flight secondary ion mass spectrometer (TOF-SIMS), probing the first few top monolayers of the samples with a pulsed primary $In^+$ beam of about 35x50 µm² (Kissel *et al.* 2007). Prior studies of 67P dust particles with COSIMA have shown that the average carbon abundance of 67P particles is close to the solar value (Bardyn et al. 2017) and that carbon is mainly in the form of high-molecular-weight organic matter (Fray *et al.* 2016). However, Mg, Fe, and other metals such as Ca and Al were depleted relative to Si at the top surface of the 67P particles compared to the chondrite reference, but only average values for a small set of samples were considered (Bardyn et al. 2017, Stenzel *et al.* 2017). In this study, we present the elemental ratios (relative to Si) for H, C, Na, Mg, Al, K, Ca and Fe from 63 surface analyses of 59 individual particles. In addition, during the last few weeks of the Rosetta mission, COSIMA performed 12 analyses of 8 cometary particles after a "sputtering" sequence with a continuous beam of primary $In^+$ ions. For each of these particles, the sputtering step removed about 250 nm of the sample's surface, revealing its subsurface composition (see Section 2.1).

# 2 METHODS

## 2.1 COSIMA instrument: collection and analysis of dust particles

COSIMA contained 72 movable collection targets, an optical microscope (COSISCOPE), and a Time-Of-Flight Secondary Ion Mass Spectrometer (TOF-SIMS) (Kissel et al. 2007). Dust particles ejected from the 67P nucleus were collected, imaged, and analysed by COSIMA to measure their composition (Hilchenbach *et al.* 2016). The one centimetre-square collection targets were exposed at the base of a funnel open to the cometary atmosphere. These targets were grouped by 3 on 24 target holders, and 7 of these holders were exposed and used for dust capture (see **Table 1**). Most of the targets were covered by a highly porous layer of gold "black" to increase the probability of dust collection (Hornung *et al.* 2014). The dust impact velocities on targets, ranging from 0.3 to 12.2 m $s^{-1}$ (Della Corte *et al.* 2015), rule out any chemical alteration of the composition of dust particles during their capture (Hornung *et al.* 2016). The internal optical camera of COSIMA (COSISCOPE), with a spatial resolution of 14 µm, enabled the assessment of the morphology and size characterisation of collected dust particles (Schulz *et al.* 2015, Hilchenbach et al. 2016, Hornung et al. 2016, Langevin et al. 2016). COSISCOPE images were also used to determine the locations of collected dust particles before measuring their chemical composition with the TOF-SIMS. A primary 8 keV $^{115}In^+$ ion beam was pulsed at a frequency of 1.5 kHz, with about 1000 ions per 3 ns pulse. The primary ion beam had a footprint area of 35 x 50 µm² (full width at half-maximum) on the target (Kissel *et al.* 2007, Hilchenbach *et al.* 2016). The impacts of the primary ions on the samples produced neutral and charged secondary fragments from the topmost atomic layers of the studied samples. Either negative or positive secondary ions could be measured, depending on the chosen polarity of the acceleration voltage. In this paper, only the positive secondary ion mass spectra were considered. Secondary ions were then separated in a reflectron TOF system. The time-of-flight of the ions was measured and converted to mass-to-charge (m/z) ratios to generate mass spectra. The reflectron improves the instrument's spectral resolution by compensating for differences in the ions' initial kinetic energy. COSIMA mass resolution was about $m/\Delta m$ = 1400 (full width at half-maximum) at m/z=100 (Hilchenbach et al. 2016), thus enabling distinguishing, in most cases, elemental ions from hydrogen-bearing organic ions for a given integer mass. During the 2-year mission, about 35,000 particles or particle fragments were collected (Merouane et al. 2017), of which 250 were analysed, yielding almost 35,000 mass spectra.

During the mission, an “analysis” was defined as a sequence of spectra acquired at a given date of the mission on a given location on the target, either as a line, a cross or a matrix over a dust particle, or on a blank target area to get a background signal. Some particles were analysed multiple times at several dates.

Before routine analyses with laboratory TOF-SIMS instruments, a “sputtering” phase is common to clean the sample surface. However, as it would use a large quantity of indium, this step was skipped in most of the COSIMA analyses to optimise the use of the limited amount of primary indium ions. Thus, most particles were analysed without sputtering. At the end of the mission, eight particles were analysed before and after sputtering. During the sputtering step, the particles were irradiated with a continuous, defocused primary indium ion beam, which eroded their surfaces before analysis (see below). We call “no-sputtering” analyses those that do not involve any sputtering step. On the contrary, we call “pre-sputtering” or “post-sputtering” analyses those that involve a sputtering step followed by subsequent analysis. Most of the mass spectra measured during the mission are “no-sputtering” analyses, with acquisition times of 2.5 or 5 minutes. At the end of the mission (summer 2016), telemetry data from Rosetta to Earth were reduced, so fewer spectra were measured, but with longer acquisition times of 15 to 20 minutes and involved a “pre-sputtering” step followed by “post-sputtering” analyses. For the sake of comparison, in this work, all spectra have been normalised to an acquisition time of 2.5 min, equivalent to 225,000 primary indium ion pulses.

During sputtering, the intensity of the continuous indium ion beam reaching the targets’ surface was about 10 nA. Its footprint on the samples was a 100 µm diameter circular spot (Kissel et al. 2007). Thus, the ion flux was about $8 \times 10^{14}$ ions $cm^{-2}\ s^{-1}$. Considering a sputtering yield of 1 and a canonical value of $2 \times 10^{15}$ atoms $cm^{-2}$ for one monolayer, the sputtering step was eroding about 0.4 monolayers $s^{-1}$. Since the sputtering step lasted 42 min, the thickness of the material sputtered from the sample's surface is estimated to be about 250 nm.

**Table 1:** Targets information and values of the selection criteria.

| Target holder | Collection timeframe YYYY/MM/DD | Target name | Target surface | Analysis type | $12^+/73^+$ | $23^+/73^+$ | $24^+/73^+$ | $28^+/73^+$ |
|---|---|---|---|---|---|---|---|---|
| D0 | 2014/11/08 to 2014/12/12 | 1D0 | Gold black 20-30 µm thick | no-sputt. | 0.01 | 0.004 | 0.0007 | 0.2 |
| | | 2D0 | Gold black 20-30 µm thick | no-sputt. | 0.006 | 0.004 | 0.0007 | 0.2 |
| | | 3D0 | Gold black 20-30 µm thick + Ag part. | no-sputt. | 0.007 | 0.002 | 0.0005 | 0.2 |
| CF | 2014/12/16 to 2015/02/09 | 1CF | Gold black 8 µm thick | no-sputt. | 0.007 | 0.003 | 0.0004 | 0.2 |
| | | 2CF | Gold black 12 µm thick + Ag part. | no-sputt. | 0.01 | 0.005 | 0.0007 | 0.2 |
| | | 3CF | Gold black 20-30 µm thick | no-sputt. | 0.008 | 0.003 | 0.0002 | 0.2 |
| C7 | 2015/02/14 to 2015/04/06 | 1C7[a] | Silver blank | no-sputt. | -- | -- | -- | -- |
| | | 2C7[a] | Silver black 21 µm thick | no-sputt. | -- | -- | -- | -- |
| | | 3C7[b] | Gold black 15 µm thick | no-sputt. | 0.008 | 0.002 | 0.0003 | 0.2 |
| D1 | 2015/04/10 to 2015/05/27 | 1D1[a] | Silver blank | no-sputt. | -- | -- | -- | -- |
| | | 2D1[c] | Gold black 13 µm thick | no-sputt. | 0.02 | 0.002 | 0.0008 | 0.3 |
| | | | | pre-sputt. | 0.01 | 0.0007 | 0.0002 | 0.2 |
| | | | | post-sputt. | 0.02 | 0.03 | 0.005 | 0.2 |
| | | 3D1 | Gold black 13 µm thick | no-sputt. | 0.03 | 0.01 | 0.001 | 0.3 |
| CD | 2015/05/30 to 2015/10/07 | 1CD | Gold black 5-8 µm thick | no-sputt. | 0.03 | 0.025 | 0.005 | 0.2 |
| | | 2CD | Gold black 14 µm thick | no-sputt. | 0.02 | 0.005 | 0.0008 | 0.3 |
| | | 3CD | Gold black 20-30 µm thick | no-sputt. | 0.03 | 0.005 | 0.002 | 0.3 |
| D2 | 2015/10/11 to 2016/05/08 | 1D2 | Gold black 8 µm thick | no-sputt. | 0.02 | 0.01 | 0.001 | 0.3 |
| | | 2D2 | Gold black 8 µm thick | no-sputt. | 0.02 | 0.004 | 0.0005 | 0.3 |
| | | 3D2[a] | Silver black 30 µm thick | no-sputt. | -- | -- | -- | -- |
| C3 | 2016/05/10 to 2016/09/30 | 1C3[c] | Gold black8 µm thick | no-sputt. | 0.02 | 0.03 | 0.002 | 0.3 |
| | | | | pre-sputt. | 0.02 | 0.002 | 0.0003 | 0.2 |
| | | | | post-sputt. | 0.02 | 0.02 | 0.002 | 0.2 |
| | | 2C3[c] | Gold black 15 µm thick | no-sputt. | 0.02 | 0.05 | 0.0007 | 0.2 |
| | | | | pre-sputt. | 0.02 | 0.05 | 0.0007 | 0.2 |
| | | | | post-sputt. | 0.06 | 0.04 | 0.007 | 0.4 |
| | | 3C3 | Gold black 20-30 µm thick | no-sputt. | 0.02 | 0.006 | 0.0004 | 0.2 |

[a] The silver targets have not been considered in this work as they are less clean than the gold ones.

[b] The selection criteria of 3C7 target only apply for particles analysed between 2015/11/12 and 2015/11/13, as technical problems have occurred during spectra acquisition at the other dates.

[c] Sputtering has been performed on particles from these targets. Thus, selection criteria are divided into three types: no-sputtering, pre-sputtering, and post-sputtering.

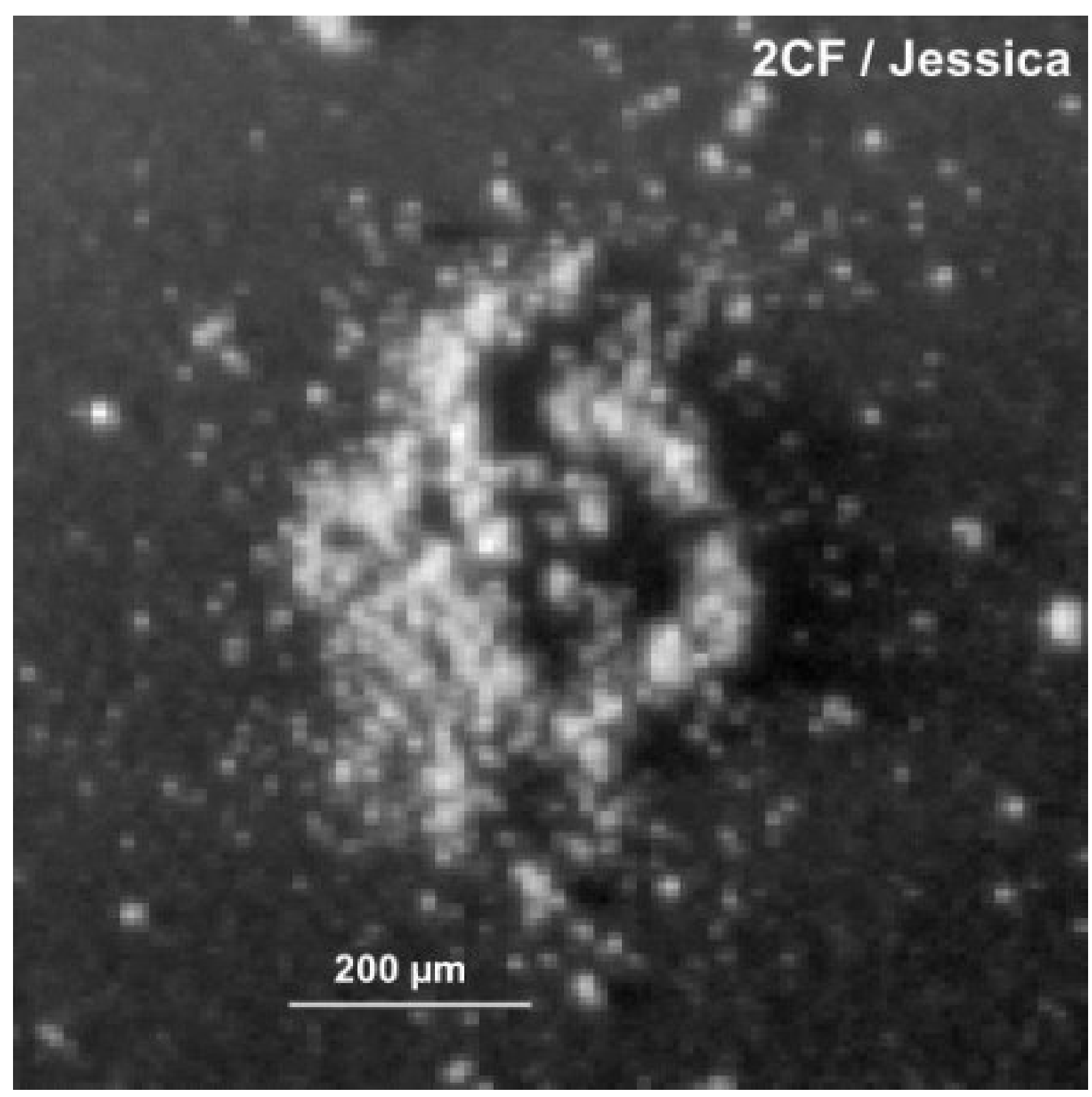

**Figure 1.** COSISCOPE image of the 2CF/Jessica particle. The light is coming from the left and the spatial resolution is about 10 µm pixel$^{-1}$ (Langevin et al. 2016).

When collected by the COSIMA instrument, most dust particles were fragmented, either by collisions with the entrance funnel or by impact with the sampling target (Langevin et al. 2016, Merouane et al. 2017) (see **Figure 1**). These fragments have sizes ranging up to a few hundred micrometres on the target. It has been shown that they are agglomerates of units with sizes of tens of µm that were unfragmented during the collection (Hornung et al. 2016, Merouane et al. 2017, Hornung *et al.* 2023), which themselves are agglomerates of smaller individual subunits, possibly down to scales of ~50 nanometers as revealed by the Rosetta atomic force microscope MIDAS (Mannel *et al.* 2019) (see **Figure 2**). As result of this fragmentation upon collection, COSIMA did not analyse the surface of the individual dust particles as ejected from the nucleus of 67P, but instead probed the surface of the fragments of the shattered particles. Consequently, the analyses conducted after sputtering the particles provided information on the composition of the fragments' subsurface.

Particle fragments analysed with COSIMA have received a name (see **Table 2**). The analyses by mass spectrometry were carried out at least one week after the samples were collected. Between collection and analysis, the particles resided inside the COSIMA instrument, where the temperature is about 300 K (Kissel et al. 2007). Thus, the particles contained no ice at the time of their analysis, which focuses solely on the most refractory fraction of the cometary particles.

**Table 2**: List of the particles analyses used is this paper and ions counts for elements. Only a few lines and rows of the table are shown as an example, the full table is available in an online table.

| N° | Target | Particle (First Name) | Particle (Last Name) | Starting date of the collection period | Ending date of the collection period) | Date of the Analysis | Number of considered spectra | H+ counts | ±σ | C+ counts | ±σ | Na+ counts | ±σ |
|---|---|---|---|---|---|---|---|---|---|---|---|---|---|
| *No-sputtering analyses* | | | | | | | | | | | | | |
| 1 | 1D0 | Sigrid(1) | Vesijako | 14/11/2014 | 21/11/2014 | 21/04/2016 | 50 | 8,06E+02 | 7,29E+00 | 1,19E+02 | 2,22E+00 | 6,95E+02 | 4,02E+00 |
| 2 | 2D0 | Estelle-Alizée(1) | Nilakka | 25/09/2014 | 03/10/2014 | 27/04/2016 | 56 | 5,19E+02 | 9,75E+00 | 6,68E+01 | 2,20E+00 | 2,31E+02 | 2,52E+00 |
| 3 | | Estelle-Alizée(2) | Nilakka | 25/09/2014 | 03/10/2014 | 28/04/2016 | 4 | 9,89E+01 | 5,05E+00 | 9,34E+00 | 1,56E+00 | 3,76E+01 | 3,08E+00 |
| 4 | 3D0 | Marius-Elias | Saimaa | 25/10/2014 | 31/10/2014 | 06/05/2016 | 6 | 5,40E+02 | 1,73E+01 | 1,14E+02 | 6,18E+00 | 2,44E+02 | 6,79E+00 |
| 5 | 1CF | Cleo | Ala-Kitka | 16/12/2014 | 20/12/2014 | 18/06/2016 | 1 | 7,00E+02 | 2,70E+01 | 1,05E+02 | 8,92E+00 | 9,07E+01 | 7,60E+00 |
| 6 | | Daniel | Kolima.3 | 24/01/2015 | 25/01/2015 | 17/06/2016 | 1 | 4,00E+02 | 2,17E+01 | 7,10E+01 | 8,31E+00 | 1,15E+02 | 8,64E+00 |
| 7 | | Erkka | Enonvesi | 20/12/2014 | 27/12/2014 | 18/06/2016 | 3 | 8,25E+02 | 2,76E+01 | 1,26E+02 | 7,76E+00 | 3,24E+02 | 8,40E+00 |
| 8 | | Fanny | Kolima.3 | 24/01/2015 | 25/01/2015 | 18/06/2016 | 1 | 4,45E+02 | 2,76E+01 | 7,85E+01 | 9,14E+00 | 1,09E+02 | 8,39E+00 |
| 9 | | Françoise | Ala-Kitka | 16/12/2014 | 20/12/2014 | 18/06/2016 | 1 | 5,98E+02 | 4,20E+01 | 1,24E+02 | 1,32E+01 | 5,84E+02 | 1,86E+01 |
| 10 | | Gervas | Ala-Kitka | 16/12/2014 | 20/12/2014 | 18/06/2016 | 2 | 6,74E+02 | 3,05E+01 | 1,11E+02 | 9,09E+00 | 2,80E+02 | 9,57E+00 |
| 11 | | Glenn | Lummene.4 | 28/01/2015 | 29/01/2015 | 17/06/2016 | 1 | 5,55E+02 | 2,46E+01 | 7,49E+01 | 8,96E+00 | 1,28E+02 | 8,96E+00 |
| 12 | | Jan | Kolima.3 | 24/01/2015 | 25/01/2015 | 18/06/2016 | 1 | 8,12E+02 | 3,77E+01 | 1,58E+02 | 1,21E+01 | 1,89E+02 | 1,08E+01 |
| 13 | | Joar | Lummene.4 | 28/01/2015 | 29/01/2015 | 18/06/2016 | 1 | 1,02E+03 | 4,41E+01 | 1,78E+02 | 1,39E+01 | 2,97E+02 | 1,36E+01 |
| 14 | | Katharina | Kolima.3 | 24/01/2015 | 25/01/2015 | 18/06/2016 | 2 | 5,29E+02 | 2,40E+01 | 8,57E+01 | 6,86E+00 | 1,03E+02 | 5,82E+00 |
| 15 | | Kenzi | Ala-Kitka | 16/12/2014 | 20/12/2014 | 17/06/2016 | 1 | 5,22E+02 | 2,26E+01 | 9,24E+01 | 9,19E+00 | 1,76E+02 | 1,02E+01 |
| 16 | | Khady | Ala-Kitka | 16/12/2014 | 20/12/2014 | 18/06/2016 | 1 | 1,35E+03 | 4,22E+01 | 2,14E+02 | 1,32E+01 | 4,40E+02 | 1,58E+01 |
| 17 | | Klaus | Inari | 02/01/2015 | 09/01/2015 | 18/06/2016 | 3 | 3,59E+02 | 1,19E+01 | 6,68E+01 | 4,44E+00 | 2,63E+01 | 2,66E+00 |
| 18 | | Lenz | Lummene.1 | 25/01/2015 | 26/01/2015 | 18/06/2016 | 2 | 1,27E+03 | 3,22E+01 | 1,88E+02 | 9,86E+00 | 3,40E+02 | 1,02E+01 |
| 19 | | Modou | Ala-Kitka | 16/12/2014 | 20/12/2014 | 18/06/2016 | 2 | 4,74E+02 | 1,74E+01 | 8,45E+01 | 5,96E+00 | 1,32E+02 | 6,20E+00 |
| 20 | | Silivia | Kolima.3 | 24/01/2015 | 25/01/2015 | 17/06/2016 | 2 | 5,17E+02 | 2,65E+01 | 8,13E+01 | 7,71E+00 | 8,05E+01 | 5,67E+00 |
| 21 | | Stavro | Kolima.3 | 24/01/2015 | 25/01/2015 | 17/06/2016 | 1 | 3,78E+02 | 2,74E+01 | 6,85E+01 | 8,96E+00 | 1,01E+02 | 8,15E+00 |

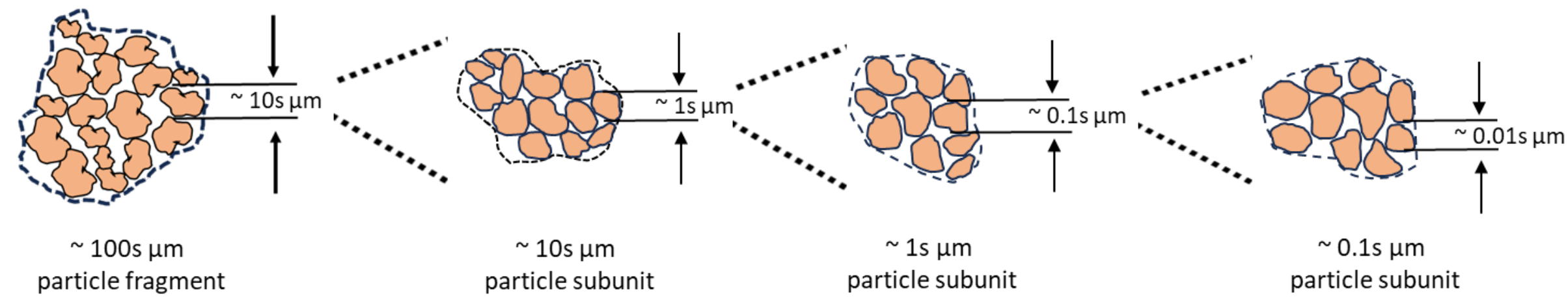

**Figure 2**. Schematic sketch of the hierarchical model of the porous cometary dust particles.

## 2.2 Search for mass spectra of interest

During the 2-year mission, 21 targets from 7 target holders were exposed to the cometary environment. Of these 21 collection targets, 17 were gold foils covered with porous gold black, and 4 were made of silver (see **Table 1**). We observed that silver targets had a high level of surface contamination, so they were not used to extract relevant particle composition data. Only mass spectra measured on the gold targets were considered. In the positive ion mass spectra of the targets before exposure to the cometary atmosphere, or after collection, PolyDiMethylSiloxane (PDMS) polymers $(-(CH_3)_2-Si-O)_n$ can be detected from their characteristic fragments at m/z=73.05 ($Si(CH_3)_3^+$) and m/z=147.07 $(Si_2O(CH_3)_5)^+$. Originating from electronic components, PDMS is expected in the background of any TOF-SIMS mass spectra due to its very high ionisation yield (Vickerman & Briggs 2001, Henkel & Gilmour 2014). The $^{197}Au^+$ ions at m/z=196.97 are also detected, originating from the gold target material, as well as a signature of the implanted $^{115}In^+$ primary ions at m/z=114.90. All the spectra performed on the targets themselves, i.e. before dust collection or at locations where no cometary particles are visible on the COSISCOPE images, are called “background spectra”. These “background spectra” are selected and analysed before the mass spectra of cometary particles. This first step allows us to characterise the background signal on each target. In the background signal, the most intense peak is the $Si(CH_3)_3^+$ ion before sputtering, if no sputtering is implemented, and $^{115}In^+$ after sputtering (see **Figure 3**).

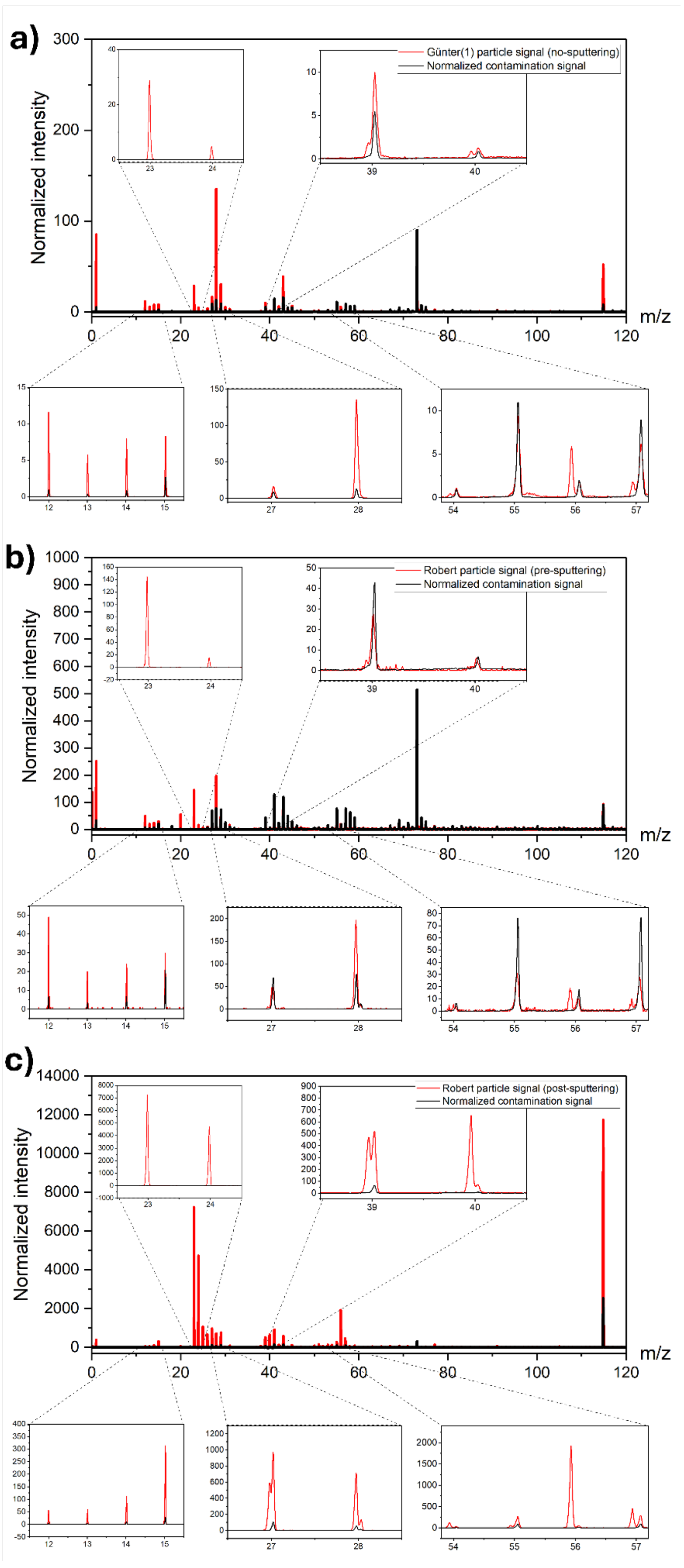
a)
Normalized intensity
m/z
Günter(1) particle signal (no-sputtering)
Normalized contamination signal
b)
Robert particle signal (pre-sputtering)
Normalized contamination signal
c)
Robert particle signal (post-sputtering)
Normalized contamination signal

**Figure 3**. Comparison of mass spectra performed before and after sputtering. The different panels present the average mass spectrum of the no-sputtering analysis of Günter(1) (panel a), pre-sputtering analysis of Robert particle (panel b) and post-sputtering analysis of Robert particle (panel c). For these three analyses, the average mass spectrum of the particles is shown in red, and the associated background spectra are shown in black. All average spectra obtained from particles were normalised to an integration time equivalent to 2 min 30, corresponding to 225,000 counts, whereas the associated background spectrum was normalised to the PDMS intensity. Inserts highlight peaks in different mass ranges.

Previous studies (Hilchenbach et al. 2016, Bardyn et al. 2017) have shown that the mass spectra measured on cometary particles contain both signals from cometary particles and from the background (**Figure 3**). However, the detection of elemental ions such as $^{23}Na^+$ (m/z= 22.99), $^{24}Mg^+$ (m/z=23.99) and $^{56}Fe^+$ (m/z= 55.94), is characteristic of the cometary signal (Hilchenbach et al. 2016, Bardyn et al. 2017). Moreover, even if $^{12}C^+$ and $^{28}Si^+$ ions are also detected in the background signal, the ionic ratios $^{12}C^+/Si(CH_3)_3^+$ and $^{28}Si^+/Si(CH_3)_3^+$ are much higher in the cometary signal than in the background signal (**Figure 3**).

To select the mass spectra with a significant cometary signal, a systematic procedure based on the main differences between cometary and background signals was used. First, the total number of counts at each integer mass was calculated for all the spectra. At m/z = 28, the signal is mainly due to $^{28}Si^+$, even if a small contribution of $C_2H_4^+$ is also observed in all spectra. At m/z = 12, 23, and 24, the signal is only due to $C^+$, $Na^+$ and $Mg^+$, whereas m/z = 73 is characteristic of PDMS only. The background spectra were defined as those with more than 100 counts at m/z=73 and for which the $12^+/73^+$, $23^+/73^+$, $24^+/73^+$, and $28^+/73^+$ ionic ratios are below appropriate thresholds (see **Table 1**). These thresholds are established based on the mass spectra acquired on the targets before dust collection. On the other hand, the "cometary" spectra were defined as those for which the $12^+/73^+$, $23^+/73^+$, $24^+/73^+$ and $28^+/73^+$ ionic ratios are higher than the same thresholds and with more than 10 counts at m/z=23. Based on the criteria for the number of counts at m/z 23 and 73, only mass spectra with a sufficient signal-to-noise ratio were considered. Different thresholds were chosen for the "no-sputtering", "pre-sputtering", and "post-sputtering" mass spectra. This methodology led to the selection of 63 "no-sputtering," 8 "pre-sputtering," and 12 "post-sputtering" analyses for this study. The 8 "pre-sputtering" and the 12 "post-sputtering" analyses correspond to 8 different cometary particles (some of them have been analysed more than once after sputtering). All the analyses used in this paper are listed in **Table 2**. Finally, all the mass spectra of a given particle on a specific date, i.e., belonging to the same analysis, were added together and normalised to an acquisition time of 2.5 minutes, yielding an average spectrum. Likewise, an average background spectrum was calculated by summing all "background" spectra obtained from a specific target area and normalised to an acquisition time of 2.5 min. In this paper, we call "local background" the "background" spectra acquired at the same date and at the vicinity of the corresponding "particles" spectra. Some typical averaged spectra for "no-sputtering", "pre-sputtering", and "post-sputtering" analyses are presented in **Figure 3**.

## 2.3 Cometary signal extraction

Positive ion mass spectra were first mass-calibrated using the COSIMA pipeline. Then, the peaks attributed to the elemental ions of interest were fitted with Gaussian-like profiles (a sum of three Gaussian curves) using a Levenberg-Marquardt algorithm. For all fitted peaks, the centroid, width, and counts number were retrieved, with a peak fit residual of generally about 0.3 percent. The uncertainty in the count numbers accounts for the systematic error of the fit and the statistical error (**Table 2**). To determine the number of counts of all elemental ions of interest, this adjustment procedure was applied to the normalised sum of all mass spectra measured for a given analysis, as well as to the associated local background spectra, if it is available, and to an averaged background of each collection target. To retrieve the particle's signal in the spectra, the contribution of the background signal, mainly due to PDMS, is subtracted using Eq. (1):

$$X^+_{corrected} = X^+_{uncorrected} - X^+_{background} \times f_{norm} \quad (1)$$

In Eq. (1), $X^+_{uncorrected}$ is the number of counts of a specific ion in a spectrum measured on a particle, $X^+_{background}$ is the number of counts of the same ion in the background spectrum and $X^+_{corrected}$ is the corrected number of counts. In this equation, all the number of counts are normalized to an acquisition time of 2.5 min. $f_{norm}$ is a normalisation factor calculated with Eq. (2)

$$f_{norm} = \frac{1}{2} \times \left( \frac{[Si(CH_3)_3^+]_{sample}}{[Si(CH_3)_3^+]_{background}} + \frac{[Si_2O(CH_3)_5^+]_{sample}}{[Si_2O(CH_3)_5^+]_{background}} \right) \quad (2)$$

Where, $Si(CH_3)_3^+$ and $Si_2O(CH_3)_5^+$ are the number of counts on PDMS fragments at m/z=73.05 and 147.07, respectively.

The abundances of nine elemental ions were quantified in this work: $H^+$, $C^+$, $Na^+$, $Mg^+$, $Al^+$, $Si^+$, $K^+$, $Ca^+$ and $Fe^+$. As $H^+$, $C^+$, $Na^+$, $Mg^+$, $Si^+$, and $Fe^+$ are detected in most of the background mass spectra, the subtraction procedure was applied to these ions. On the contrary, no subtraction was needed for $Al^+$, $K^+$ and $Ca^+$, as they are not detected in the background mass spectra. We have found that using a local background or an average background yielded the same results (see APPENDIX A). Thus, only the correction with the average background will be considered further (see APPENDIX B).

## 2.4 Particle Analysis Selection

We focused on the abundances of 9 elements (H, C, Na, Mg, Al, K, Ca and Fe) in the spectra. $H^+$, $C^+$, $Na^+$, $Mg^+$, $Si^+$ and $Fe^+$ were generally detected with a good signal-to-noise ratio in the "no sputtering" analyses. On the contrary, the intensities of $Al^+$, $K^+$ and $Ca^+$ peaks were low, and they are located next to intense peaks of hydrocarbon ions with long tails toward lower m/z ratios. Thus, the deconvolution of $Al^+$, $K^+$ and $Ca^+$ peaks was sometimes complex and/or led to high uncertainties in the determination of the number of counts in the selected peaks. We thus only considered ion peaks with a relative uncertainty on their total number of counts that is lower than 33% (at 1 $\sigma$) (**Table 2**). Therefore, the number of analyses in which the elements can be quantified can vary from one element to another.

## 2.5 Quantification of the elemental composition of cometary matter

Elemental ratios can be deduced from measured ionic ratios using calibration factors known as RSF (Relative Sensitivity Factors). They have been calculated from calibration mass spectra acquired with the reference model (RM) of COSIMA located in Göttingen (Germany) on reference standard samples for which elemental ratios X/Y are known (Krüger *et al.* 2015).

$$RSF(X/Y) = \frac{(X^+/Y^+)_{std}}{(X/Y)_{std}} \quad (3)$$

In Eq. (3), X and Y are two elements, $(X^+/Y^+)_{std}$ is the ionic ratio measured in the mass spectra on standard samples and $(X/Y)_{std}$ is the known elemental ratio of the reference standard sample.

For a given elemental ratio, the corresponding RSF may depend on the nature of the sample being studied. This effect is generally called the “matrix effect” and constitutes a common bias in TOF-SIMS analysis. Taking this effect into account is essential, but it is also very challenging. Mineral standards of various types, such as pyroxenes, olivines, feldspars, hydrated silicates, and carbonates, have been analysed using the COSIMA RM (Krüger et al. 2015). As RSFs are not necessarily the same from one mineral to another, average RSF(X/Y) values for Na, Mg, Al, Si, K, Ca and Fe have been calculated from these standards, with significant uncertainties reflecting the matrix effect (Krüger et al. 2015). It is also worth noting that adding a sputtering phase on these calibration samples did not significantly change the RSFs (Krüger et al. 2015). The RSFs used in this study are shown in **Table 3**. The RSFs used in this study to quantify Mg and Fe differ from those used in prior works (Bardyn et al. 2017). Indeed, for consistency and robustness, we have chosen to use only calibrations made with the ground RM COSIMA model on a large set of mineral samples, rather than a single acquisition on an olivine standard collected with the flight model. This explains the shift toward lower Mg/Si and Fe/Si values in this study compared to our previously published average value.

**Table 3:** RSF values used for quantification before and after sputtering in positive mode. All the RSF values are taken from Krüger *et al.* (2015), except the ones for H/C and C/Si which have been calculated by Bardyn *et al.* (2017) and Isnard *et al.* (2019). These values come from the analysis of a series of samples with the Reference Model (RM) of COSIMA. They are averages of measurements taken on various samples (see Krüger *et al.* (2015)), except the one for C/Si which is derived from the analysis of a single CSi sample, also with RM instrument.

| Element | RSF value | σ+ | σ- |
|---|---|---|---|
| **H/C** | $0.21(H^+/C^+)-0.38$ | 1.10E-01 | 1.10E-01 |
| **C/Si** | 2.50E-02 | 7.00E-03 | 5.00E-03 |
| **Na/Si** | 3.50E+01 | 6.70E+01 | 2.30E+01 |
| **Mg/Si** | 3.70E+00 | 2.20E+00 | 1.40E+00 |
| **Al/Si** | 3.50E+00 | 3.20E+00 | 1.70E+00 |
| **K/Si** | 6.40E+01 | 6.20E+01 | 3.10E+01 |
| **Ca/Si** | 7.60E+00 | 4.30E+00 | 2.70E+00 |
| **Fe/Si** | 1.80E+00 | 1.10E+00 | 7.00E-01 |

# 3 RESULTS

**Figure 4** shows the elemental composition of the surface (without or before sputtering) and of the subsurface (after sputtering) of 67P dust particle fragments normalised to Si. These compositions are compared to those of the solar photosphere, the CI chondrite reference, 1P/Halley, and 81P/Wild 2 cometary dust. CI chondrites have a solar composition for most elements and are depleted in volatile elements, such as H, C, N, and noble gases. Based on this comparison, the composition of 67P dust particles reveals several key features:

(i) Except for H and C, all the measured elemental abundances relative to Si at the surface of the 67P particle fragments are lower than in the Sun and in the CI chondrite. For Mg, Ca, and Fe, depletions by at least a factor of 10 are observed in 67P dust and can reach more than a factor of 1000 for some particle fragments.

(ii) The surface and the subsurface of 67P dust particle fragments have different compositions. In subsurface measurements, Na, Mg, Al, K, Ca, and Fe relative to Si are more abundant compared to the surface composition. However, H/Si and C/Si remain roughly similar. On average, the composition of the subsurface is chondritic relative to Si for Mg, Al, K, Ca, and Fe.

(iii) The C/Si elemental ratios for both surface and subsurface analyses are larger than in CI chondrites and close to the solar value, which confirms the previously reported high abundance of carbonaceous phases in 67P particles (Bardyn *et al.* 2017). In contrast, all H/Si values are close to the chondritic value. Regarding Na, on average, its subsurface elemental ratio to Si is above the chondritic reference value, as measured in bulk for particles of comet 81P/Wild 2 (Stephan 2008) (but unlike 1P/Halley's value, which has a chondritic value (Jessberger et al. 1988)).

(iv) Elemental ratios from subsurface measurements of 67P fragmented dust particles roughly encompass the average values measured for 1P/Halley and 81P/Wild 2 dust particles (Jessberger et al. 1988, Stephan 2008). Except for the lighter elements (H, C and Na), the elemental abundances relative to Si in the 1P/Halley particles are lower than in the Sun and the CI chondrite reference. Even if the observed depletions are much larger at the surface of the 67P particle fragments than in the 1P/Halley dust, the dust from both comets shares similarities.

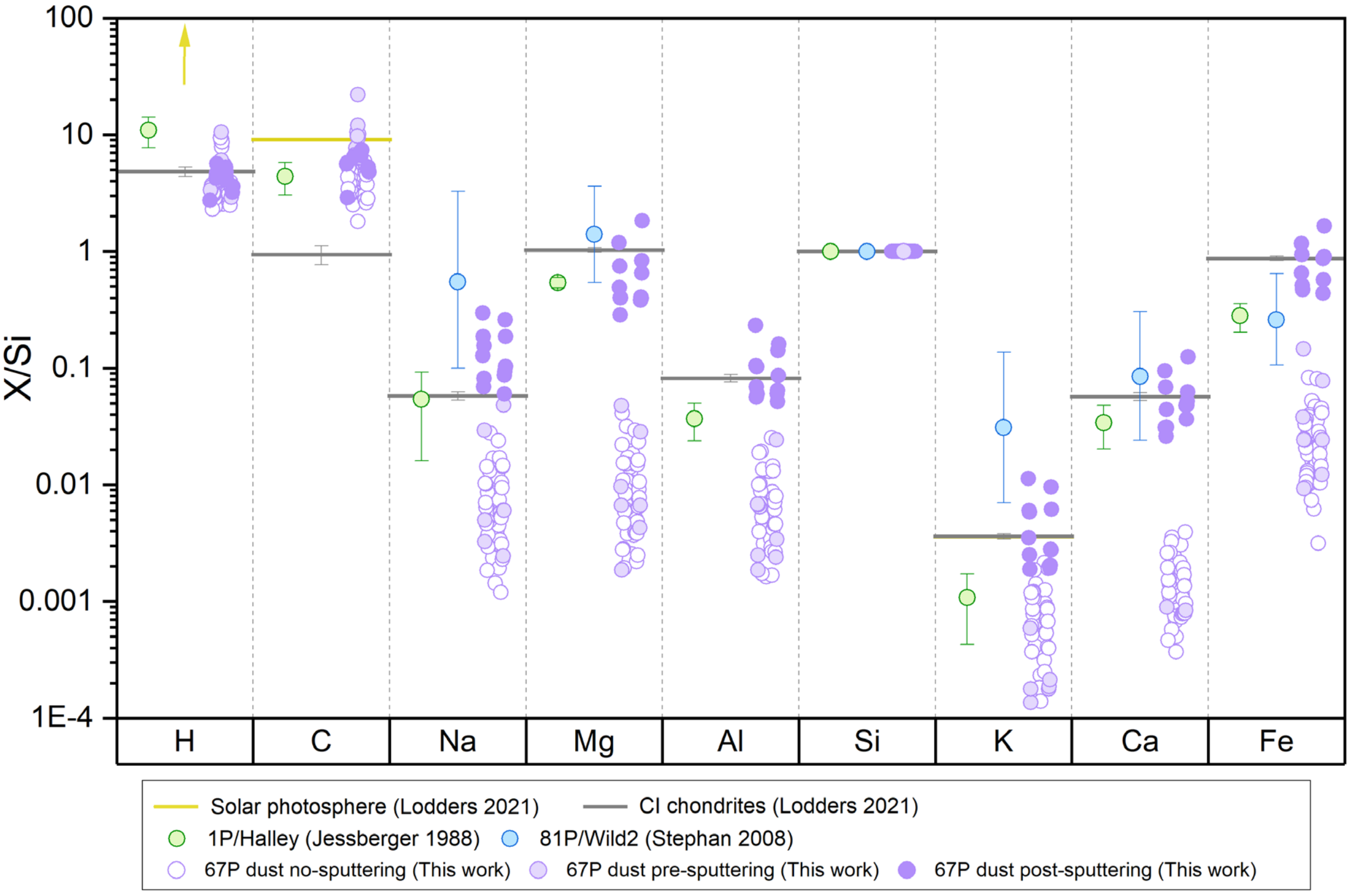

**Figure 4.** Elemental composition of comet 67P/Churyumov-Gerasimenko fragmented dust particles normalised to Si. 63 analyses of 59 different particle fragments without sputtering (purple open dots), of 8 particles before sputtering (plain light purple dots), and of 12 particle fragments' subsurface analyses after sputtering (plain dark purple dots). Data for the solar photosphere (yellow horizontal lines or arrow for H) (Lodders 2021), CI chondrites (horizontal grey lines) (Lodders 2021), comet 1P/Halley (green dots) (Jessberger *et al.* 1988), and comet 81P/Wild2 (blue dots) are also plotted for reference. For 81P/Wild 2, the geometric means (and associated geometric standard deviations) were calculated from measurements performed on samples collected on Al foils (Stephan 2008), which allows normalisation to Si. The solar and CI compositions are indistinguishable in this plot for elements other than H and C. For 67P dust particles, only elemental ratios with relative errors lower than 33% are shown (see Methods and **Table 2**). To improve the clarity of this figure, error bars for 67P data are not shown in this figure. Still, they are displayed for Mg and Fe in **Figure 5** and **Figure B1**, and **Figure D1** and **Figure D2** for all elements.

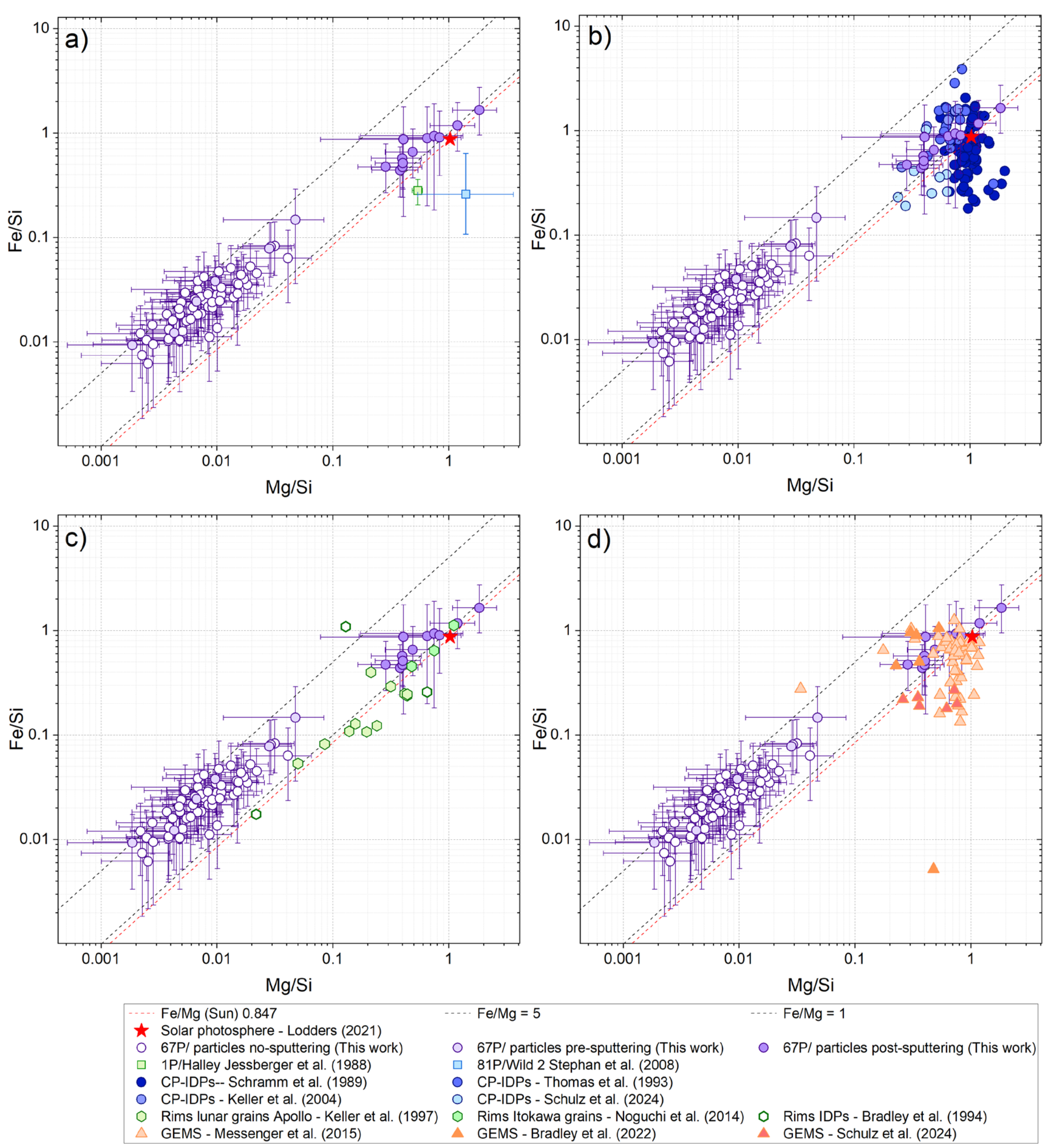

**Figure 5.** Fe/Si as a function of Mg/Si for comet 67P dust particle fragments, as measured by COSIMA. Fe/Si and Mg/Si elemental ratios measured at the surface (without or before sputtering – purple open dots and light purple dots) and subsurface (after-sputtering – dark purple dots) of 67P dust particle fragments, compared to: comets 1P/Halley (Jessberger et al. 1988) and 81P/Wild2 (Stephan 2008) dust average values in panel a), CP-IDPs (Schramm *et al.* 1989, Thomas *et al.* 1993, Keller *et al.* 2004, Schulz *et al.* 2024) in panel b), rims observed in lunar regolith, Itokawa asteroid grains and IDPs (Bradley 1994, Keller & McKay 1997, Noguchi *et al.* 2014) in panel c) and GEMS (Bradley *et al.* 2022, Messenger *et al.* 2015, Schulz et al. 2024) in panel d). These elemental ratios are also compared to the solar photosphere value in all panels (Lodders 2021).

**Figure 5** shows Fe/Si as a function of Mg/Si for 67P particle fragments compared to values measured in relevant interplanetary objects. Fe/Si and Mg/Si values are correlated for both surface and subsurface measurements (Figure 5a). One can note, however, that the Fe/Mg value is slightly higher at the surface than at the subsurface of 67P particle fragments.

There is a very good agreement between subsurface values of 67P dust particles and the bulk composition of Chondritic Porous Interplanetary Dust Particles (CP-IDPs), considered to have a possible cometary origin (Keller et al. 2004, Thomas et al. 1993, Schramm et al. 1989, Schulz et al. 2024) (Figure 5b).

Amorphous rims are commonly observed around mineral samples collected at the surface of airless bodies, such as the Moon or asteroids (Bradley 1994, Keller & McKay 1997, Noguchi *et al.* 2014). These rims are altered amorphous layers, typically 10–100 nm thick, formed around particles exposed to space weathering (a combination of charged particle irradiation and micrometeorite bombardment) (Bennett *et al.* 2013, Pieters & Noble 2016). The Mg/Si and Fe/Si ratios in these rims are generally lower than in the core of space-weathered minerals (see Appendix C and figures therein). It has been shown experimentally that the irradiation of Mg- and/or Fe-bearing crystalline silicates by $H^+$ and $He^+$ with energies of about 1 to 10 keV (to simulate the solar wind) can induce the formation of an amorphous rim, with a decrease of the Mg/Si, and in some cases, Fe/Si ratios at the surface of the irradiated samples (Carrez *et al.* 2002, Jäger *et al.* 2016, Laczniak *et al.* 2021). Figure 5c shows the Mg/Si and Fe/Si elemental ratios measured in rims of Itokawa, lunar and IDP grains with cores made of Mg- and Fe-bearing silicates (Bradley 1994, Keller & McKay 1997, Noguchi *et al.* 2014). Among the lunar grains studied by Keller & McKay (1997), we selected those with compositions that could be relevant to that of cometary matter (see Appendix C). The limited depletion of these ratios in Itokawa grains relative to lunar samples may reflect a shorter exposure age (Noguchi *et al.* 2014, Keller *et al.* 2021). For most lunar, IDP, and Itokawa rims, the Mg/Si ratio decreases progressively from the interior to the surface of the grains (Bradley 1994, Keller & McKay 1997, Noguchi *et al.* 2014). Thus, the very low Mg/Si and Fe/Si values observed at the surface of the 67P particle fragments could be due in part to the fact that COSIMA probes only the outermost monolayers of the sample's surface.

Besides, the compositions measured at the surface of 67P dust are not comparable to those measured in GEMS (Glass with Embedded Metal and Sulfides), which are glassy phases of possible presolar origin found in CP-IDPs (Messenger *et al*. 2015, Bradley *et al*. 2022) (Figure 5d).

In addition, no dependence is observed between the surface or subsurface elemental ratios of the analysed particles and the heliocentric distance of comet 67P at the time of collection (ranging from1.26 au near perihelion to over 3.6 au at the end of the mission; see Appendix D , **Figure D1** and **Figure D2**). This shows that the elemental composition of the particles and the extent of their metal depletions are independent of cometary activity or solar flux variations during the orbit.

# 4 DISCUSSION

Rims observed in lunar regolith samples and asteroid returned samples are explained by surface space weathering processes on the respective objects. Indeed, the surfaces of these airless bodies are continuously bombarded by micrometeorites and charged energetic particles from the solar wind. These processes alter the structure and the chemical composition of the outermost layer of the grains exposed to them. They lead to a depletion in Mg and Fe, which is a key feature of the rims in lunar regolith, Itokawa grains and some IDPs. As shown in this work, the surface of 67P particle fragments is also depleted in Mg and Fe compared to their subsurface. Therefore, we conclude that the cometary particle fragments analysed by COSIMA, or their subunits, are surrounded by rim-like structures formed by processes like space weathering, which occurs on airless bodies. In lunar and Itokawa samples, some minor elements, such as Ca, K, or Al, can be enriched in the rims by micrometeorite impacts (Noguchi *et al*. 2014). Such enrichments are not observed on the 67P particles (see Appendix C). We propose that the formation of Mg- and Fe-depleted rims in 67P samples is due to ion irradiation only.

To better understand the mechanisms driving these compositional gradients, laboratory experiments have investigated the effects of various forms of irradiation and bombardment on the surfaces of different sample types relevant to astrophysical objects. Recent results from ion irradiation experiments at 1 keV/amu or higher show that the formation of these rims is the result of a complex combination of different processes that include sputtering, redeposition, radiation-enhanced diffusion and segregation (Laczniak *et al.* 2021, Laczniak *et al.* 2024). The combination of these mechanisms accounts for the complex structure observed at the nanometre scale in the returned samples, even though sputtering was long considered the primary mechanism explaining elemental depletions in irradiated layers. Sputtering itself is affected by several parameters, such as the element's mass, its abundance in the sample, and its surface binding energy. In experiments where silicates such as olivine are irradiated by 1 to ~10 keV/amu, the element preferentially sputtered is oxygen (Demyk *et al.* 2001, Carrez *et al.* 2002). Depending on the composition of the starting silicate, this slight depletion in O causes chemical rearrangements of the irradiated layer, leading either to the reduction of iron and the formation of iron nanoparticles, or to a segregation of Mg and Si resulting in an Mg- and Fe-rich outer layer above a Si-rich layer below (Laczniak *et al.* 2021, Laczniak *et al.* 2024). A first type of experiment, applied to amorphous $MgFeSiO_4$ samples irradiated with 10 to 20 keV protons, relevant to interstellar medium dust processing, reproduces the composition of GEMS in terms of Mg, Fe and Si abundances and mineralogy (Jäger *et al.* 2016). A second type of experiment, applied to a Murchison chondrite sample irradiated at different fluxes and fluences of 1 keV $H^+$ and 4 keV $He^+$, shows altered structures in the rims surrounding olivine grains that are very similar to those observed in detailed analyses of Ryugu and Itokawa space-weathered grains (Laczniak *et al.* 2024).

Mg and Fe depletions are observed for all the surface analyses of comet 67P particle fragments (**Figure 4** & **Figure 5**). If the depletion by irradiation took place after the agglomeration of subunits into dust particles, the samples' surface analysed by COSIMA (after fragmentation during collection) should be mainly pristine (i.e. un-irradiated, see **Figure 6**, hypothesis 1). This first hypothesis is inconsistent with the COSIMA measurements. On the contrary, if the irradiation took place on the subunits before their agglomeration into dust particles, the surface of the analysed sample, i.e. after the fragmentation of the original particle upon collection, is then entirely composed of irradiated metal-depleted subunits (see

**Figure 6**, hypothesis 2), as measured with COSIMA. Therefore, we conclude that rim formation due to irradiation occurred on the mineral subunits before they agglomerated into aggregates, thereby building up the cometary dust particles. This is consistent with analyses of CP-IDPs (Bradley 1994) and one ultra-carbonaceous Antarctic micrometeorite (UCAMM) (Engrand *et al.* 2020) in which rims are observed around minerals inside the particles, suggesting irradiation of the mineral constituents happened before their incorporation into the CP-IDPs or UCAMMs.

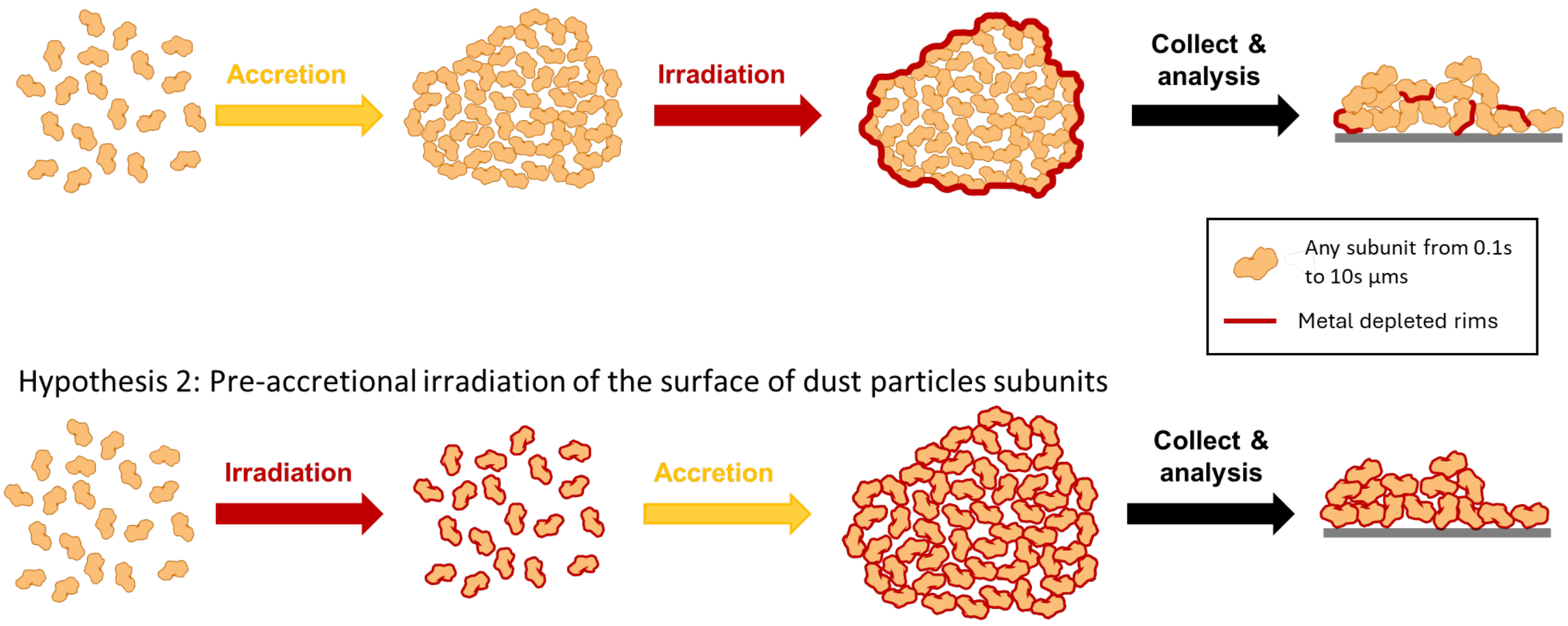

**Figure 6.** Schematic representation of the accretion of 67P dust particles and their analysis with COSIMA, according to the timeline of accretion and irradiation processes of the initial constituents and rim formation. Orange constituents are subunits of cometary dust particles analysed by COSIMA. Red plain lines represent the metal-depleted rims. Considering the thickness of the surface removed by sputtering and the typical thickness of rims, rough geometrical considerations suggest that space weathering and rim formation could have occurred on subunits with sizes ranging from hundreds of nanometers to tens of micrometers, while still preserving a pristine core.

COSIMA analysed the composition of the surface of agglomerates of unfragmented units with sizes of tens of µm (Hornung et al. 2016, Merouane et al. 2017, Hornung et al. 2023). Rim thicknesses are variable both in extraterrestrial samples (Bradley 1994, Keller & McKay 1997, Noguchi et al. 2014) and samples irradiated in the laboratory (Demyk *et al.* 2001, Carrez *et al.* 2002, Jäger *et al.* 2016, Laczniak *et al.* 2021), ranging from a few tens to about a hundred nanometres at the surface of an irradiated material. This rim thickness is mainly controlled by the energy and dose of the impinging ions(Poppe *et al.* 2018), with a maximum thickness of 100-200 nm in lunar or asteroidal samples. Therefore, COSIMA analyses after $In^+$ sputtering (removing about 250 nm of the sample surface, see Section 2.1) probe the composition of particles below any rim layer, revealing the inner composition of 67P cometary dust (see **Figure 7**).

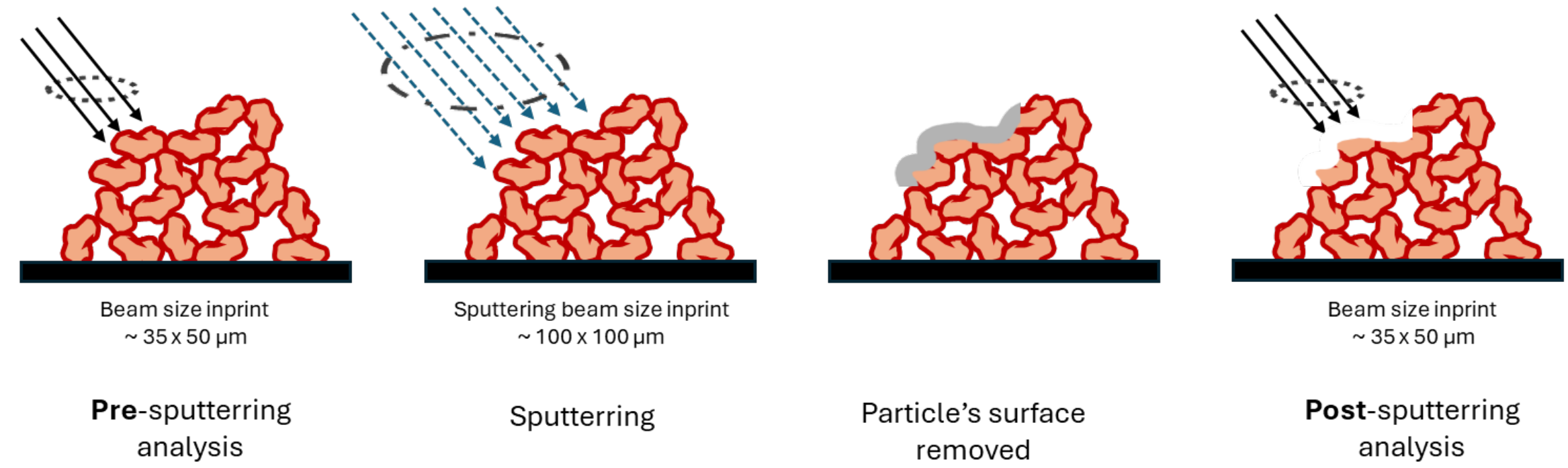

**Figure 7.** Sketch of the analytical sequence from the "pre-sputtering" analysis to the "post-sputtering analysis. The "post-sputtering" analyses probe the dust particles composition below the irradiation rims. For clarity, in this example, the size of the irradiated subunits is a few tens of micrometres, but it would also be consistent at the micrometre or a few hundred nanometre levels (see main text). The thickness of the rims and the depth of the removed surface are not at scale.

The formation of the rims on minerals before their accretion in cometary particles could have occurred either in the interstellar medium or in the solar nebula. Only a specific range of ion energy induces changes in composition for irradiated minerals, usually accompanied by amorphization of the material (Demyk *et al*. 2001, Carrez *et al*. 2002, Jäger *et al*. 2016). Indeed, energies below 1 keV do not lead to significant mineral alteration, while ions above 50 keV penetrate deep into minerals without sputtering the surface elements. These energy considerations rule out irradiation of cometary material by high-energy protons, which dominate galactic cosmic rays (above 1 MeV) (Thoudam *et al.* 2016). However, in supernovae shockwaves, $H^+$ and $He^+$ have the appropriate energies to interact with the surface of crystalline silicate grains formed near red giant stars and alter their composition (Demyk *et al*. 2001, Molster & Kemper 2005). In a paradigm where comets would be inherited from a pristine interstellar composition (Greenberg 1998), this could explain observations presented in this work. However, this scenario suffers a limitation as the silicate grains' size distribution in number in the diffuse interstellar medium follows a power law with grain radius between 25 and 250 nm and an exponent between -3.3 and -3.6 (Mathis *et al.* 1977). Therefore, interstellar dust is dominated, in number and mass, by the smallest grains. Because these small interstellar grains are in the same size range as the rims formed by irradiation weathering, such a process would have impacted most, if not all, of their volume. Thus, this interstellar scenario does not appear to be consistent with our measurements, which show a clear difference between the surface and the subsurface of the cometary particles.

In the solar nebula, solar wind particles are dominated by protons, with a smaller contribution of $He^+$ and heavier ions with ~ 1 keV / nucleon (Gosling 2007), i.e., within an efficient energy range for the preferential sputtering of Mg and other elements (Carrez et al. 2002, Laczniak et al. 2021). Thus, the formation of rims due to solar wind irradiation could have occurred in the protoplanetary disc.

The Mg/Si and Fe/Si ratios are significantly depleted in rims compared to core composition for most lunar soil particles. Therefore, while keeping in mind that this variation is not as pronounced as in 67P particles and that lunar soil is not a relevant proxy for cometary matter regarding mineralogy and composition, we presume that the ion fluence responsible for the Mg and Fe depletion at the surface of 67P particle fragments must have been higher than the one that formed the lunar rims. The exposure age of lunar grains, estimated from track densities, is about $10^6$ years (Keller *et al*. 2021). In the protoplanetary disc, it is assumed that particles with a size comparable to that collected by COSIMA (around tens or hundreds of micrometres) could accrete in just a few $10^4$ years (Weidenschilling 1997, Laibe *et al.* 2008), i.e. the accretion time of the subunits surrounded by a rim could be shorter than the exposure ages of most lunar grains by a factor of about 100. Considering that the flux of the solar wind of the young Sun was likely similar to the present one (Wood *et al.* 2014), the ion irradiation of cometary particle subunits should have taken place in the inner part of the protoplanetary disc at approximately 0.1 AU from the Sun.

An additional process could contribute to the metal depletions at the surface of 67P dust fragments. Laboratory experiments have shown that ion-irradiated silicates become highly reactive to water, leading to a rapid cation exchange that removes up to 60% of surface Mg and other weakly bonded metals (Cantando *et al.* 2008). In cometary dust, water ice mixed with irradiated mineral subunits before accretion could trigger a similar depletion process. As the ice sublimates near the Sun, water molecules would carry away loosely bound cations, leaving behind an outer rim more depleted in metals.

This scenario of irradiation in the inner part of the protoplanetary disc is supported by analyses of samples returned from comet 81P/Wild 2, which concluded that it had accreted material heavily processed by heat and intense ion flux in the inner and warmer regions of the solar nebula (Marty *et al.* 2008, Brownlee 2014). To date, no amorphous or altered rims have been reported in 81P/Wild 2 samples, but if such rims surrounded the subunits of the captured particles, the high-velocity impact in the collecting aerogel would likely have destroyed them.

The composition of the particles from comet 1P/Halley is generally intermediate when compared with those of the surfaces and sub-surfaces of the particles from comet 67P (see **Figure 4**). This could be explained by the fact that the compositions of the comet 1P/Halley particles were measured as a whole, after high-speed impacts and ionisation on a metallic surface. Those of the 67P particles were obtained on a few monolayers at the surface of the aggregates, before or after sputtering. If the structure of the particles from comet 1P/Halley is similar to that of 67P, the measurements obtained for comet 1P/Halley would therefore represent an "average" incorporating the surfaces and sub-surfaces of the particles.

Thermal infrared spectra of cometary dust often reveal the presence of crystalline silicates (Harker et al. 2023). As the mineral material inherited from the interstellar medium is mostly amorphous (Kemper *et al.* 2004), the crystalline silicates in comets must have formed in the inner and warm parts of the solar nebula (Bockelée-Morvan *et al.* 2002). While modelling used to fit infrared observations of dust particles typically assumes a mixture of pure crystalline silicates with pure amorphous particles (Harker *et al.* 2023), we suggest here that these observations could also be accounted for, at least in part, by dust subunits that crystallised in the inner solar nebula and subsequently developed an amorphized outer rim.

After the rim formation, the dust particle subunits would have been transported to the outer solar nebula, beyond the snow line, where they aggregated into larger particles, such as those analysed with COSIMA, and ultimately into the 67P nucleus. The mineral subunits, with a surface depleted in Mg and Fe, should have a typical size ranging from a hundred nanometres (given the rim thickness ranging from a few tens of nanometres to a hundred nanometres) to a few tens of micrometres (subunits composing the 67/ particle fragments (Hornung et al.

2023)). Based on CP-IDPs and UCAMMs observations, we favour a size of several hundred nanometres for the mineral subunits with rims.

The broad range of correlated Mg and Fe depletions in surface analyses of cometary dust, as shown in **Figure 5**, may partly reflect differences in the total irradiation fluence received by individual particle subunits. Yet, beyond this effect, the trend can also be interpreted as a mixing line between two populations of constituents: (i) constituents surrounded by metal-depleted rims, and (ii) unaltered or minimally altered constituents with near-solar composition, originating from the outer nebula or from unprocessed interstellar material. The dust of comet 67P may thus consist of mixtures of these two populations. Assuming end-member compositions as rimed subunits strongly depleted in Mg and Fe and unrimed subunits of solar composition, processed rimed components should dominate since the surface Mg/Si and Fe/Si ratios measured on 67P particle fragments are at least ten times lower than solar. Therefore, they likely account for at least 90% of the subunits, implying efficient transport and incorporation of irradiated material from the inner solar nebula into the comet-forming region. However, other data from the Rosetta spacecraft suggest a potential interstellar legacy based on the D/H ratio measured in volatile molecules (Altwegg *et al.* 2019) and refractory organic matter (Paquette *et al.* 2021). Thus, ices and refractory organic matter should have accreted with the minerals in the outer part of the protoplanetary disk after the minerals were irradiated in the inner part. A sample return from comet 67P would be essential to confirm the presence of the rims discussed in this work and determine their abundance.

# 5 Conclusion

The ESA Rosetta mission has provided a unique opportunity to study comet 67P/Churyumov–Gerasimenko, thanks to a suite of complementary scientific instruments to unravel its nature and fundamental properties, offering unprecedented information into its composition and the mechanisms at play in the protosolar nebula 4.5 Gy ago. In this work, we investigated the surface and subsurface elemental compositions of the cometary dust particles measured by the COSIMA time-of-flight secondary ion mass spectrometer aboard Rosetta spacecraft. Surface analyses of 67P particle fragments reveal that their outermost monolayers are strongly depleted in Mg, Fe, Ca, Al, Na, and K relative to Si compared to solar abundances. In contrast, post-sputtering analyses removing the top ~ 250 nm shows that the subsurface material exhibits near-solar elemental ratios. This clear contrast between metal-depleted outer rims and pristine inner cores demonstrates that the observed elemental gradients reflect a surface alteration resulting from space weathering.

We conclude that these surface depletions were driven by low-energy (~ 1 keV/nucleon) ion irradiation from the solar wind. Achieving the large Mg and Fe depletions observed on 67P requires a high ion fluence implying that the mineral subunits were irradiated in the warm, inner regions of the protoplanetary disc. The measured surface ratios indicate that these irradiated, rimed subunits account for a large fraction of the mineral mass of 67P dust, demonstrating efficient outward radial transport beyond the snow line before their final co-accretion with pristine ices and refractory organic matter into the cometary nucleus.

# Acknowledgements

COSIMA was built by a consortium led by the Max-Planck-Institut für Extraterrestrische Physik, Garching, Germany, in collaboration with: the Laboratoire de Physique et Chimie de l'Environnement et de l'Espace, Orléans, France; the Institut d'Astrophysique Spatiale, CNRS/Université Paris Sud, Orsay, France; the Finnish Meteorological Institute, Helsinki, Finland; the Universität Wuppertal, Wuppertal, Germany; von Hoerner und Sulger GmbH, Schwetzingen, Germany; the Universität der Bundeswehr, Neubiberg, Germany; the Institut für Physik, Forschungszentrum Seibersdorf, Seibersdorf, Austria; and the Institut für Weltraumforschung, Österreichische Akademie der Wissenschaften, Graz, Austria; and is led by the Max-Planck-Institut für Sonnensystemforschung, Göttingen, Germany. We acknowledge the support of the national funding agencies of Germany (DLR, grant 50 QP 1302), France (CNES), Austria (project FWF P26871-N20), Finland and the ESA Technical Directorate. Rosetta is an ESA mission with contributions from its Member States and NASA. We thank the Rosetta Science Ground Segment at ESAC, the Rosetta Mission Operations Centre at ESOC and the Rosetta Project at ESTEC for their outstanding work enabling the science return of the Rosetta Mission. I.S. was supported by a PhD grant funded by CNES and Universite Paris Est Creteil. This research was funded, in part, by l'Agence Nationale de la Recherche (ANR), project ANR-24-CE49-3347-01. Authors from French laboratories acknowledge financial support from the Centre national d'études spatiales (CNES), France (ROR: https://ror.org/04h1h0y33), within the framework of the Rosetta space mission.

This paper is dedicated to the memory of Martin Hilchenbach, PI of the COSIMA instrument since 2007, deceased during the summer of 2025.

# Data availability

The COSIMA mass spectra are available via the Rosetta archive at https://www.cosmos.esa.int/web/psa/rosetta.

# References

Altwegg K., Balsiger H., Fuselier S.A., 2019, Annual Review of Astronomy and Astrophysics, 57, 113

Bardyn A., et al., 2017, Month. Not. Roy. Astr. Soc., 469, S712

Bell J.F., Davis D.R., Hartmann W.K., Gaffey M.J., 1989, in: Binzel R.P., Gehrels T., Matthews M.S. eds., Asteroids II University of Arizona Press, p. 921

Bennett C.J., Pirim C., Orlando T.M., 2013, Chemical Reviews, 113, 9086

Bockelée-Morvan D., Gautier D., Hersant F., Huré J.-M., Robert F., 2002, Astron. Astophys., 384, 1107

Bradley J.P., 1994, Science, 265, 925

Bradley J.P., Ishii H.A., Bustillo K., Ciston J., Ogliore R., Stephan T., Brownlee D.E., Joswiak D.J., 2022, Geochimica et Cosmochimica Acta, 335, 323

Brownlee D., 2014, Ann. Rev. Earth and Plan. Sci., 42, 179205

Cantando E.D., Dukes C.A., Loeffler M.J., Baragiola R.A., 2008, Journal of Geophysical Research: Planets, 113

Carrez P., Demyk K., Cordier P., Gengembre L., Grimblot J., D'Hendecourt L., Jones A.P., Leroux H., 2002, Meteoritics and Planetary Science, 37, 1599

Clark B.C., Mason L.W., Kissel J., 1987, Astron. Astophys., 187, 779

Della Corte V., et al., 2015, Astron. Astophys., 583, A13

Demyk K., et al., 2001, Astron. Astophys., 368, L38

Engrand C., et al., 2020, Lunar and Planetary Science Conference. Houston, United States,

Engrand C., Lasue J., Wooden D.H., Zolensky M.E., 2024, in: Meech K.J., Combi M.R., Bockelée-Morvan D., Raymond S.N., Zolensky M.E. eds., Comets III. p. 577

Flynn G.J., Nittler L.R., Engrand C., 2016, Elements, 12, 177

Fray N., et al., 2016, Nature, 538, 72

Gosling J.T., 2007, in: McFadden L.-A., Weissman P.R., Johnson T.V. eds., Encyclopedia of the Solar System (Second Edition). Academic Press, San Diego, p. 99

Greenberg J.M., 1998, Astron. Astophys., 330, 375

Hanner M.S., Bradley J.P., 2004, in: Festou M.C., Keller H.U., Weaver H.A. eds., Comets II. University of Arizona Press, p. 555

Harker D.E., Wooden D.H., Kelley M.S.P., Woodward C.E., 2023, The Planetary Science Journal, 4, 242

Henkel T., Gilmour J., 2014, in: Holland H.D., Turekian K.K. eds., Treatise on Geochemistry (Second Edition). Elsevier, Oxford, p. 411

Hilchenbach M., et al., 2016, Astrophys. J. Lett., 816, L32

Hornung K., et al., 2014, Planet. Space Sci., 103, 309

Hornung K., et al., 2023, Planet. Space Sci., 236, 105747

Hornung K., et al., 2016, Planet. Space Sci., 133, 63

Isnard R., et al., 2019, Astron. Astophys., 630, A27

Jäger C., Sabri T., Wendler E., Henning T., 2016, Astrophys. J., 831, 66

Jessberger E., 1999, Space Science Reviews, 90, 91

Jessberger E.K., Christoforidis A., Kissel J., 1988, Nature, 332, 691

Keller L.P., Berger E.L., Zhang S., Christoffersen R., 2021, Meteoritics & Planetary Science, 56, 1685

Keller L.P., McKay D.S., 1997, Geochimica et Cosmochimica Acta, 61, 2331

Keller L.P., Messenger S., Flynn G.J., Clemett S., Wirick S., Jacobsen C., 2004, Geochimica Et Cosmochimica Acta, 68, 2577
Kemper F., Vriend W.J., Tielens A.G.G.M., 2004, Astrophys. J., 609, 826
Kissel J., et al., 2007, Space Science Reviews, 128, 823
Kissel J., et al., 1986a, Nature, 321, 336
Kissel J., et al., 1986b, Nature, 321, 280
Kolokolova L., Hanner M.S., Levasseur-Regourd A.-C., Gustafson B.Å.S., 2004, in: Festou M.C., Keller H.U., Weaver H.A. eds., Comets II. University of Arizona Press, p. 577
Kolokolova L., Kelley M.S.P., Kimura H., Hoang T., 2024, in: Meech K.J., Combi M.R., Bockelée-Morvan D., Raymond S.N., Zolensky M.E. eds., Comets III. p. 621
Krüger H., et al., 2015, Planet. Space Sci., 117, 35
Laczniak D.L., Thompson M.S., Christoffersen R., Dukes C.A., Clemett S.J., Morris R.V., Keller L.P., 2021, Icarus, 364, 114479
Laczniak D.L., Thompson M.S., Christoffersen R., Dukes C.A., Morris R.V., Keller L.P., 2024, Icarus, 410, 115883
Laibe G., Gonzalez J.-F., Fouchet L., Maddison S.T., 2008, Astron. Astophys., 487, 265
Langevin Y., et al., 2016, Icarus, 271, 76
Lawler M.E., Brownlee D.E., 1992, Nature, 359, 810
Lodders K., 2021, Space Science Reviews, 217, 44
Mannel T., et al., 2019, Astron. Astophys., 630, A26
Marty B., et al., 2008, Science, 319, 75
Mathis J.S., Rumpl W., Nordsieck K.H., 1977, Astrophys. J., 217, 425
Merouane S., et al., 2017, Month. Not. Roy. Astr. Soc., 469, S459
Messenger S., Nakamura-Messenger K., Keller L.P., Clemett S.J., 2015, Meteoritics & Planetary Science, 50, 1468
Molster F., Kemper C., 2005, Space Science Reviews, 119, 3
Mumma M.J., Charnley S.B., 2011, Annual Review of Astronomy and Astrophysics, 49, 471
Noguchi T., et al., 2014, Meteoritics & Planetary Science, 49, 188
Paquette J.A., et al., 2021, Month. Not. Roy. Astr. Soc., 504, 4940
Pieters C.M., Noble S.K., 2016, Journal of Geophysical Research: Planets, 121, 1865
Poppe A.R., Farrell W.M., Halekas J.S., 2018, Journal of Geophysical Research: Planets, 123, 37
Sandford S.A., et al., 2010, Meteoritics and Planetary Science, 45, 406
Schramm L.S., Brownlee D.E., Wheelock M.M., 1989, Meteoritics, 24, 99
Schulz B., Vollmer C., Leitner J., Keller L.P., Ramasse Q.M., 2024, Geochimica et Cosmochimica Acta, 378, 153
Schulz R., et al., 2015, Nature, 518, 216
Stenzel O.J., et al., 2017, Month. Not. Roy. Astr. Soc., 469, S492
Stephan T., 2008, Space Science Reviews, 138, 247
Thomas K.L., Blanford G.E., Keller L.P., Klöck W., McKay D.S., 1993, Geochimica et Cosmochimica Acta, 57, 1551
Thoudam S., Rachen J.P., van Vliet A., Achterberg A., Buitink S., Falcke H., Hörandel J.R., 2016, Astron. Astophys., 595, A33
Vickerman J.C., Briggs D., 2001. IM Publication and Surface Spectra Limited,
Weidenschilling S.J., 1997, Icarus, 127, 290
Willacy K., et al., 2015, Space Science Reviews, 197, 151
Wood B.E., Müller H.-R., Redfield S., Edelman E., 2014, Astrophys. J. Lett., 781, L33

# Appendix A. Evaluation of local versus averaged background subtraction methods.

The background signal considered for each analysis of cometary dust can be taken in target areas devoid of cometary particles, but close to the analysed dust particles and at the same date ("local background"), or as an "averaged background" calculated for each target by the average of all background mass spectra, as defined in our selection procedure (see Methods). In our previous papers, only the local background was used to extract the cometary signal from mass spectra (Bardyn et al. 2017, Isnard et al. 2019). For the 63 “no-sputtering" analyses in this study, we have considered the corrected number of counts for the elemental ions of interest either by considering the “local background” or the “averaged background”. A comparison of the corrected number of counts for these 63 analyses and the 6 ions for which background subtraction is required showed that the corrected number of counts does not depend significantly on the choice of the background (local or averaged) (see **Figure A 1**). Thus, in this work, we used the averaged background to retrieve the number of counts corresponding to the cometary signal, even when the local background was available. “Pre-sputtering” and “post-sputtering” analyses of the 8 additional particles were not associated with local background. But some “background” spectra have been acquired on the same dates as the “pre-sputtering” and “post-sputtering” analyses. In this case, we have averaged these backgrounds to correct the analyses.

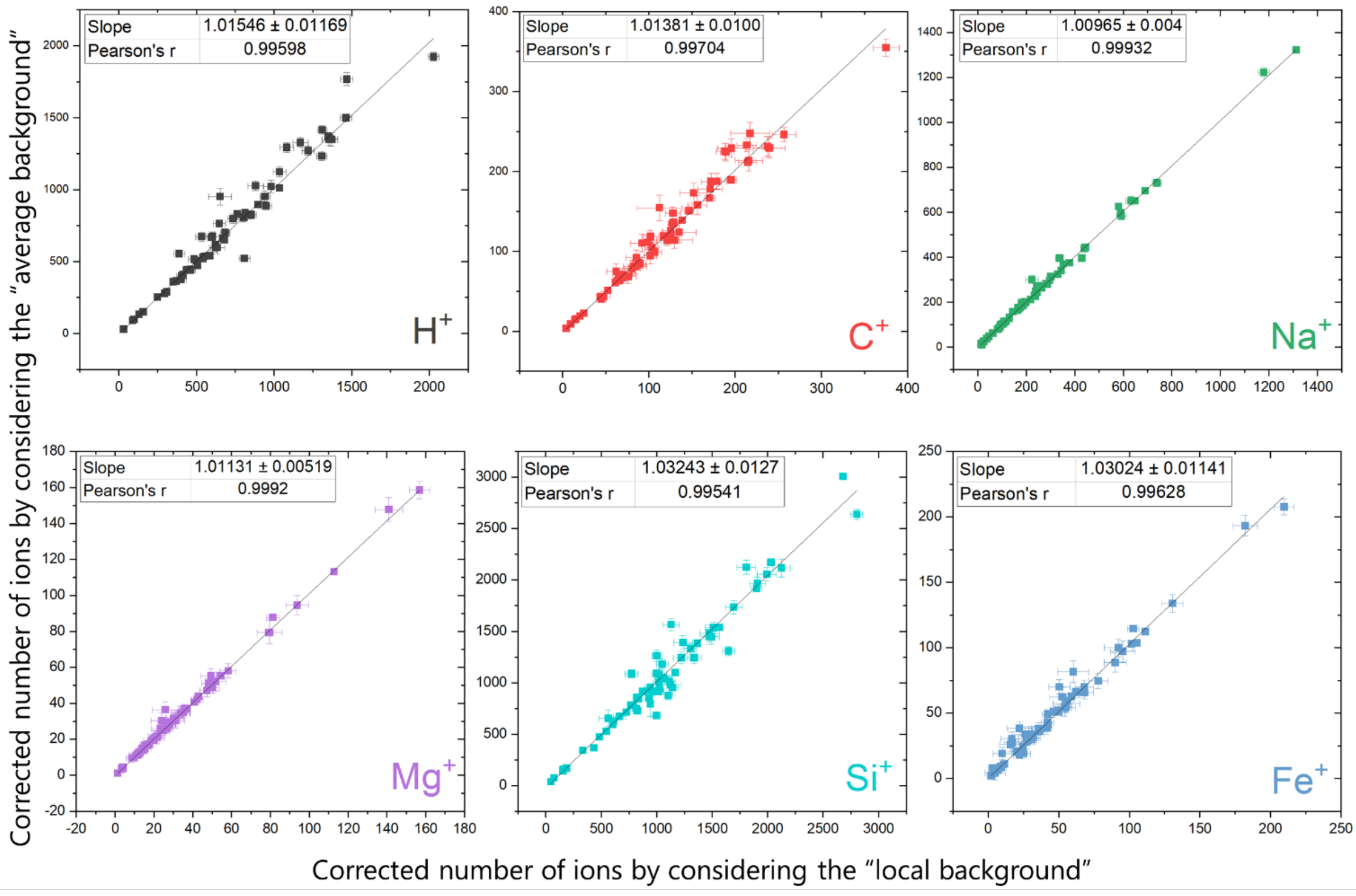

**Figure A 1**. Ion counts corrected by both types of correction. Corrected number of counts after subtraction of the “averaged background” as a function of the corrected number of counts with the “local background”. Each panel displays the number of counts for a specific elemental ion.

# Appendix B. Reliability of the background subtraction for the silicon signal.

The background signal in COSIMA is dominated by peaks characteristic of PDMS (PolyDiMethylSiloxane), a surface contaminant commonly observed in surface analyses such as TOF-SIMS (Henkel & Gilmour 2014, Vickerman & Briggs 2001). A proper quantification of the background signal, mainly due to PDMS, and its subtraction from the signal are challenging for measuring the elemental composition of 67P/Churyumov-Gerasimenko dust particles. As described in the Methods section of this paper, the averaged contamination of the collection target, onto which a particle is analysed, is subtracted from the mass spectrum measured on this given particle. However, the spectral signature of the contamination could differ on the collection targets and on the particles due to the matrix effect. Therefore, the validity of the correction can be questioned. Positive ion mass spectra of the contamination are dominated by PDMS, made of Si, O, C, and H. As Si is the element to which all the others are normalised, we focus here specifically on this element.

A contribution factor noted $f_{contrib}$ in Eq. (B1) is used to assess the amplitude of background correction.

$$f_{contrib} = \frac{X^+_{corrected}}{X^+_{uncorrected}} \quad \text{(B1)}$$

In Eq. (B1), $X^+_{uncorrected}$ is the count number of a $X^+$ ion in a spectrum measured on a particle and normalised to an acquisition time of 2.5 min. $X^+_{corrected}$ is the corrected count number after background subtraction (see Methods). The closer the $f_{contrib}$ is to 1, the less the subtraction, meaning that the initial signal of $X^+$ ion in this particles' analysis is mainly from the sample and not the background.

**Figure B 1** shows Mg/Si and Fe/Si elemental ratios as a function of $Si^+$ contribution factor for all the analyses considered in this work. The Mg/Si and Fe/Si values are not correlated to the $Si^+$ contribution factor, i.e., with the amplitude of the correction applied to the $Si^+$ ion. This supports the fact that the elemental ratios retrieved after correction are not influenced by the potential presence of residual PDMS contamination in the mass spectra measured on cometary particles. Moreover, 9 no- or pre-sputtering analyses with a $Si^+$ contribution factor higher than 0.9 (i.e. the correction applied to $Si^+$ for background subtraction is less than 10% of the initial signal) present low normalisation factors (see Methods) ranging from $4\times10^{-3}$ to $3\times10^{-2}$ and contribution factors for $Mg^+$ and $Fe^+$ are both higher than 0.97. These are among the lowest values of the normalisation factor and are indicative of a low amount of the $Si(CH_3)_3^+$ and $Si_2O(CH_3)_5)^+$ peaks characteristic of PDMS in the mass spectra collected on 67P particles. The large contribution factors for $Mg^+$ and $Fe^+$, among the highest, indicate that the contamination subtracted for these ions, as for $Si^+$, is also almost negligible. Whereas the contribution of the contamination in these very good quality analyses is low, the Mg/Si and Fe/Si elemental ratios present a quite large variation, which is the same as in the measurements presenting a lower contribution factor for $Si^+$. This suggests that there is no bias due to the initial contamination level of the analysed particles and that our methodology for subtracting background and retrieving the composition of cometary particles is robust.

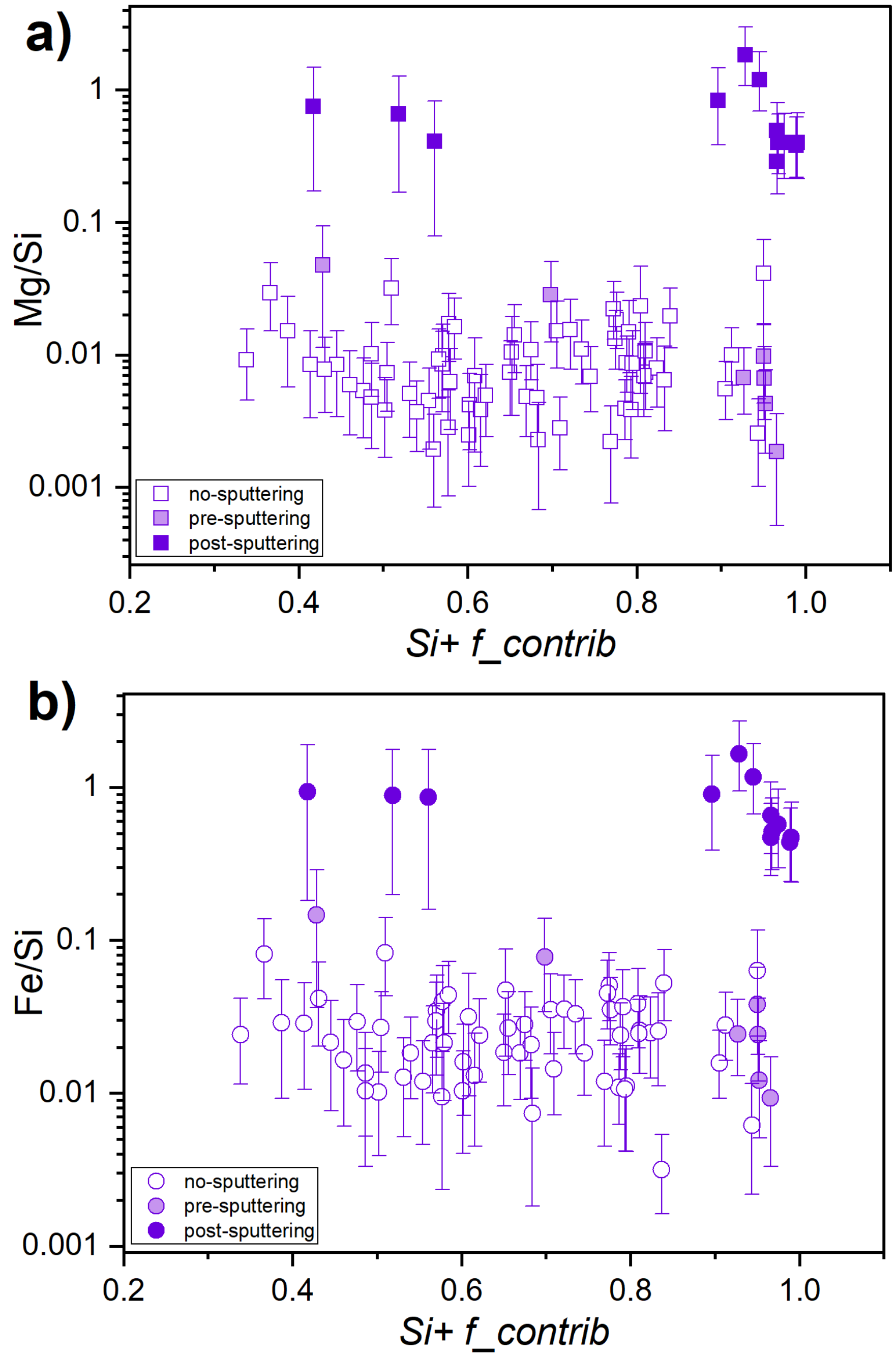

**Figure B 1**. Mg/Si and Fe/Si elemental ratios as a function of the contribution factor of $Si^+$. On each panel, no-sputtering analyses are presented in empty dots, pre-sputtering analyses in light purple filled dots and post-sputtering analyses in purple filled dots.

# Appendix C. Comparison of 67P dust composition with extraterrestrial analogues and natural rims

**Figure 4** shows the elemental composition of the surface (no sputtering), and the subsurface (after sputtering) of 67P dust particle fragments normalised to Si. Comparative data for CP-IDPs (Keller et al. 2004, Thomas et al. 1993, Schramm et al. 1989, Schulz et al. 2024), GEMS (Messenger et al. 2015, Bradley et al. 2022, Schulz et al. 2024) and natural rims on IDPs, Itokawa grains and lunar soils (Bradley 1994, Keller & McKay 1997, Noguchi et al. 2014) are presented in **Figure C 1**.

Among the lunar grains studied by Keller and McKay (1997), we selected those with compositions relevant to cometary material based on the interpretation of thermal infrared spectra of cometary dust (Harker et al. 2023). We thus selected 11 lunar grains with cores composed of clinopyroxene, orthopyroxene, and augite, and discarded grains composed of anorthite, cristobalite, and ilmenite, which should be absent or present at very low abundance in cometary dust particles. The Mg/Si and Fe/Si elemental ratios in the cores of the selected lunar sample range from 0.32 to 0.75 and from 0.16 to 0.4, respectively (Keller & McKay 1997) (see Figure C2). Based on the same criteria, we selected 3 grains of Itokawa regolith with a core made of pyroxene and olivine, but we discarded a plagioclase grain. The Mg/Si and Fe/Si values in the core of the selected Itokawa samples range from 0.63 to 1.27 and from 0.48 to 1.11, respectively (Noguchi et al. 2014) (see **Figure C2**).

The abundances of C, Na, Mg, Al, Si, Ca and Fe were measured in bulk CP-IDPs (Keller et al. 2004, Thomas et al. 1993, Schramm et al. 1989, Schulz et al. 2024). The subsurface of 67P dust particles has roughly the same average composition and dispersion values as bulk CP-IDPs, except for carbon (see Figure C1), which is more abundant in 67P dust (Bardyn et al. 2017). In contrast, GEMS display elemental ratios (Mg/Si, Al/Si, Ca/Si, and Fe/Si) that are also close to chondritic and solar values but show in most cases a narrower compositional spread than bulk CP-IDPs and match the subsurface composition of 67P dust rather than its strongly depleted surface (**Figure C 1**).

Except for H/Si and C/Si, all the elemental ratios normalised to Si present much lower values at the surface than at the subsurface of 67P dust particles (**Figure 4** and **Figure C 1**). The rims observed around lunar and Itokawa grains and around some subunits of CP-IDPs have Mg/Si and Fe/Si elemental ratios, which are almost always significantly lower than in the Sun and the CI chondrites (**Figure 4**, **Figure C 1**, **Figure C 2** and **Figure C 3**). In particular, as illustrated in **Figure C 3**, the core-to-rim pairs for lunar, asteroid, and IDP samples display in most cases a trend towards depleted Mg/Si and Fe/Si values, matching the strong surface depletion measured on 67P dust fragments.

On the contrary, the distributions of the Al/Si and Ca/Si elemental ratios measured in natural rims are approximately centred at values of the Sun and of CI chondrites (**Figure C 1** and **Figure C 2**). It has been shown that some minor elements (such as Ca, K, Al, S or Na) can be detected in the rims even if they are absent from the initial material constituting the core of these grains due to micrometeorite bombardment of the surfaces of the Moon and of Itokawa (Noguchi et al. 2014, Keller & McKay 1997). Indeed, micrometeorite bombardment can promote the formation of rims including exogenous elements from the impactor, which were not part of the initial target material. These exogenous elements could originate from minerals vaporised in the vicinity of the studied grain or simply from the impacting micrometeorites themselves (Pieters & Noble 2016). Thus, the different trends for Al/Si and Ca/Si in natural rims, compared with those in 67P particles, could indicate that the subunits of 67P particles have been subjected only to ion irradiation, and that rims have not been enriched by impacting material such as micrometeorites.

Laboratory experiments simulating the irradiation of minerals by the solar wind have shown a decrease of Mg/Si, Fe/Si and O/Si values in the samples after irradiation (Carrez et al. 2002, Demyk et al. 2001). However, the evolution of Ca and Al was not measured. Thus, irradiation experiments on silicates, during which the abundance of Ca and Al would also be followed, should be undertaken and are encouraged by the present results.

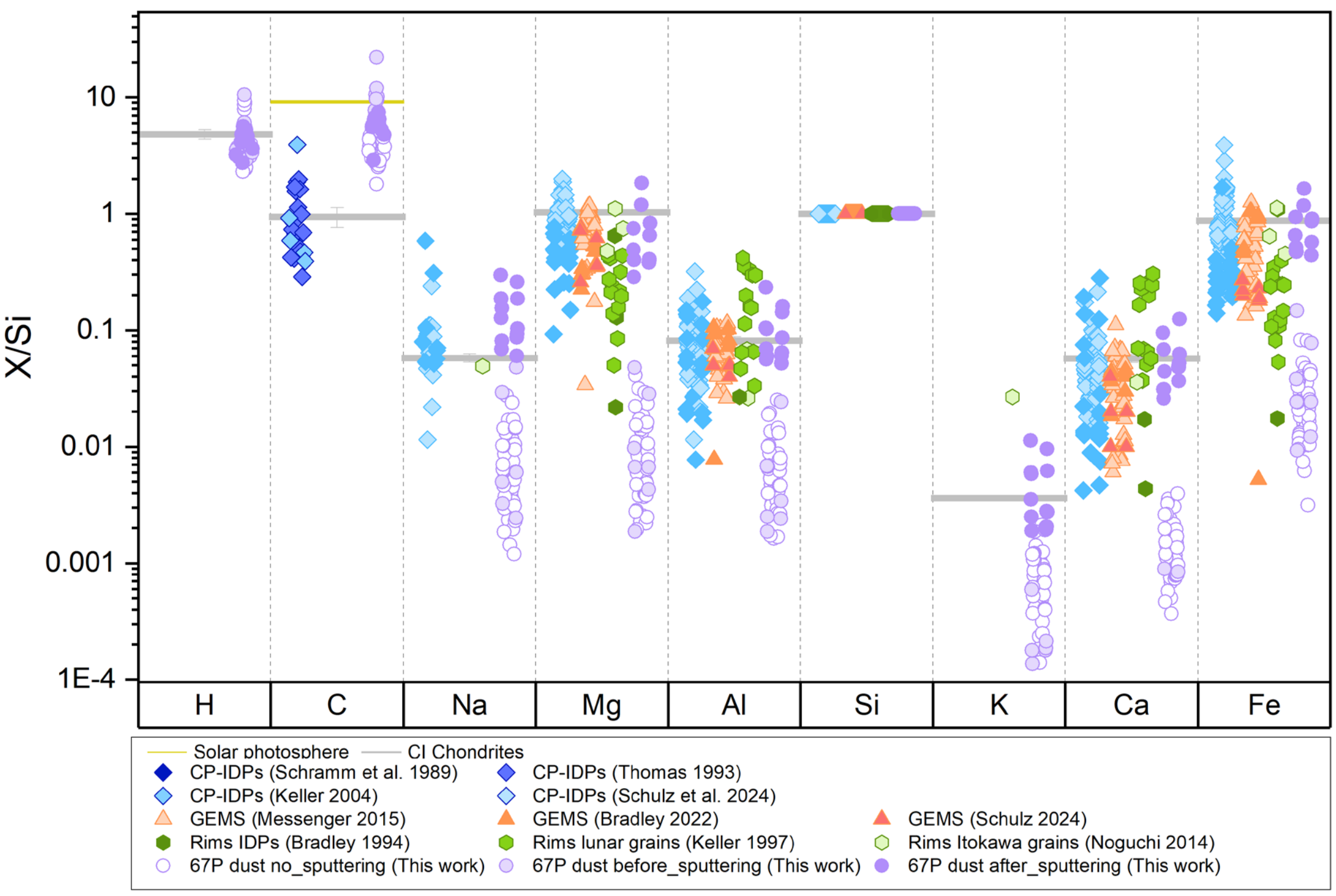

**Figure C 1.** Elemental composition of dust particles of comet 67P compared to IDPs, GEMS and natural rims. Compositions from surface analyses of 67P dust particles are presented in white (63 no-sputtering analyses) and light purple (8 pre-sputtering analyses). In contrast, sub-surface analyses are presented in purple (12 post-sputtering analyses). 67P data are compared to that of the Solar photosphere (yellow lines), CI-Chondrite (grey lines), 24 CP-IDPs (blue diamonds), 59 GEMS (orange triangles), and 19 rims from lunar regolith, IDP grains and Itokawa asteroid samples (green hexagons).

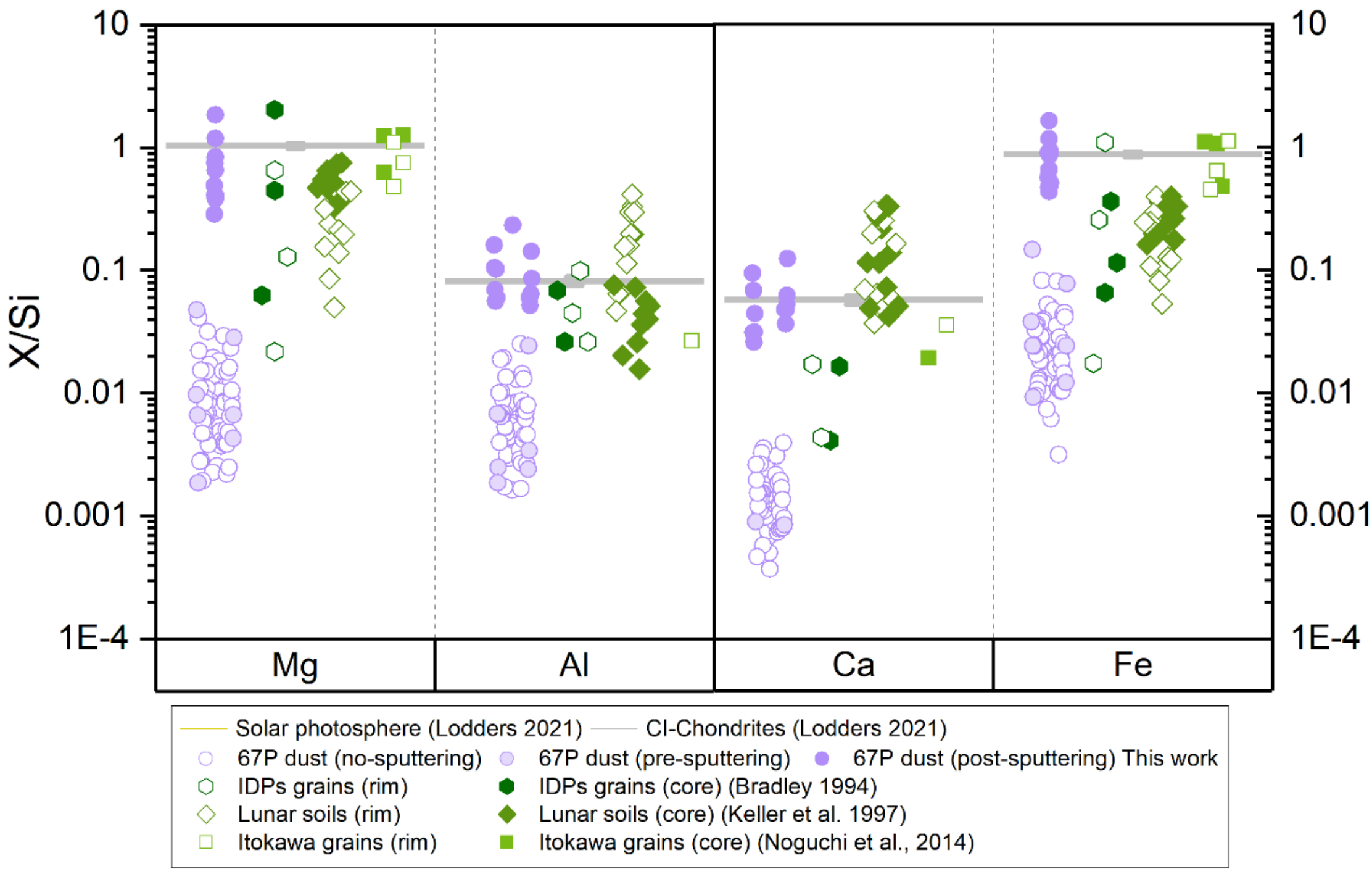

**Figure C 2**. Elemental composition of dust particles of comet 67P compared to mineral rims-bearing cores. Compositions from surface analyses of 67P dust particles are presented in white (63 no-sputtering analyses) and light purple (8 pre-sputtering analyses). In contrast, sub-surface analyses are presented in purple (12 post-sputtering analyses). 67P data are compared to that of the CI-Chondrite (grey lines) (Lodders 2021), cores and rims from lunar regolith (respectively filled and empty green diamonds), IDPs grains (filled and empty green hexagons), and Itokawa asteroid samples (filled and empty green squares).).

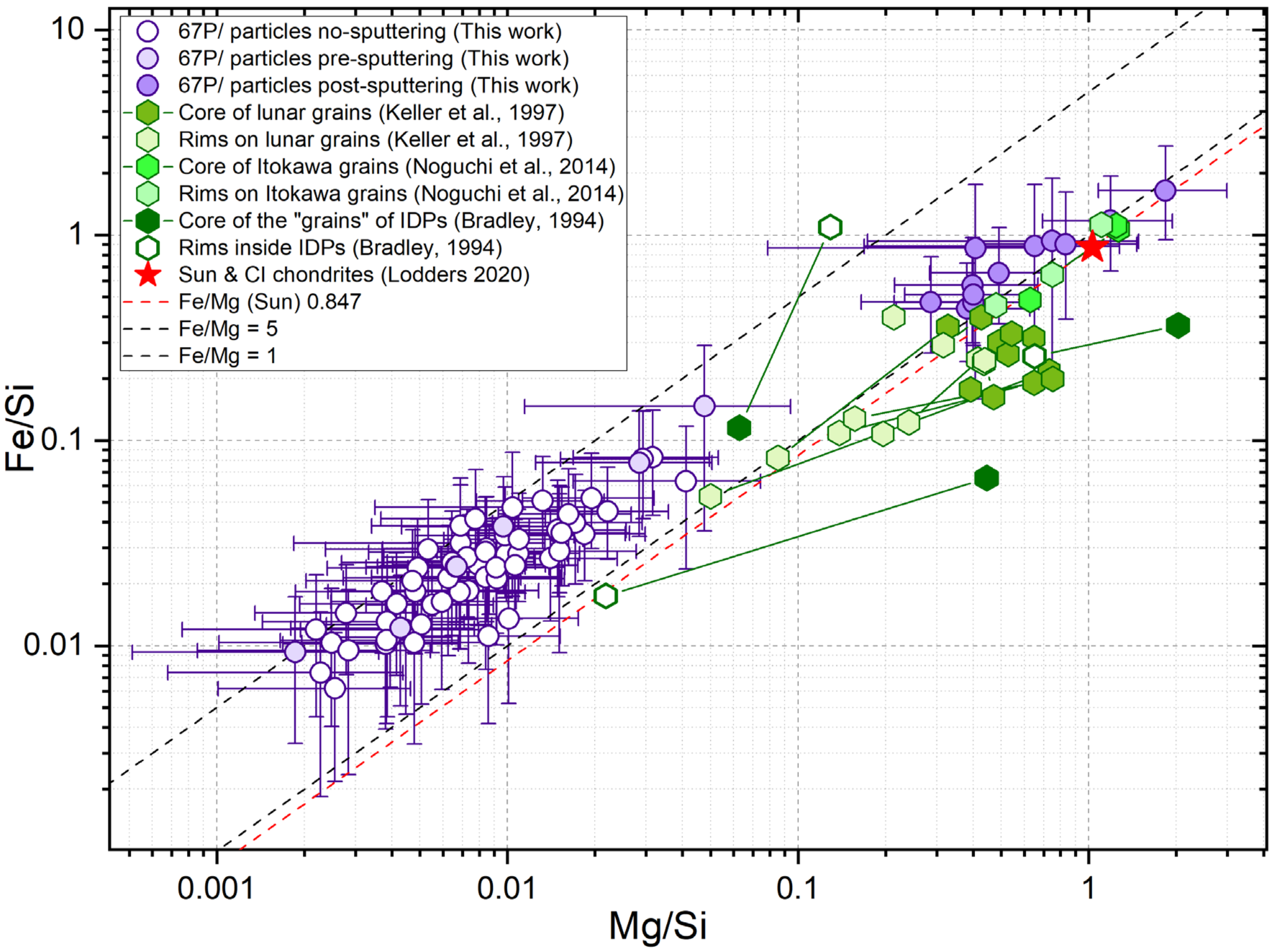

**Figure C 3.** Fe/Si and Mg/Si elemental ratios measured at the surface (without or before sputtering) and subsurface (after sputtering) of 67P dust particle fragments, compared to core and rim compositions observed in lunar regolith, Itokawa asteroid grains, and IDPs(Bradley 1994, Keller & McKay 1997, Noguchi et al. 2014). Sets of core/rim data are linked. For most of the lunar and IDPs data, a significant depletion is observed between the core and rim compositions.

## Appendix D. Composition and Heliocentric Distance

We have evaluated whether the elemental compositions measured at the surface and in the subsurface of 67P dust particles could be affected by the cometary activity along its orbit around the Sun. **Figure D 1** shows the elemental ratios relative to Si measured for surface analyses (no-sputtering and pre-sputtering) as a function of the collection date and the corresponding heliocentric distance of the nucleus, ranging from 1.26 au near perihelion (August 2015) up to about 3.6 au at the end of the mission. Similarly, Figure D2 displays the elemental ratios measured for subsurface analyses (post-sputtering) across the mission. No significant trend or correlation with heliocentric distance is observed for any of the quantified elements (H, C, Na, Mg, Al, K, Ca, and Fe relative to Si), either at the surface or in the subsurface of the particle fragments. The degree of depletion observed at the surface remains uniform across the entire orbital arc. This absence of orbital variation demonstrates that the observed surface weathering and elemental depletions are not transient effects induced by recent insolation, local thermal processing, or outgassing activity during the Rosetta mission. Instead, it confirms that the altered outer rims surrounding the mineral subunits are primordial features formed prior to the assembly of the cometary nucleus.

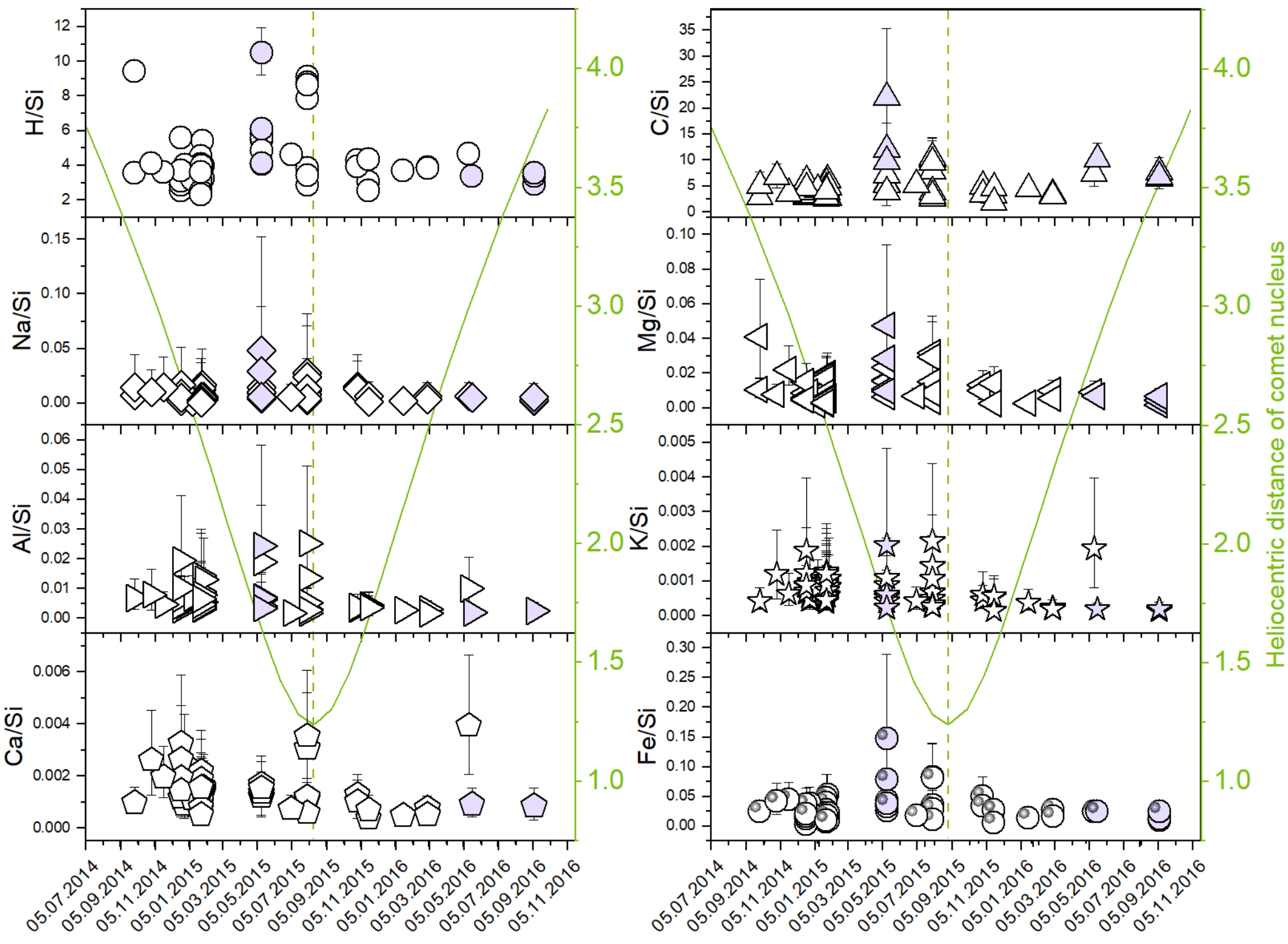

**Figure D 1.** Dust surface elemental ratios as a function of the heliocentric distance of comet nucleus at the time of the dust collection. In some cases, the error bars are not visible, as they are smaller than the symbol's size.

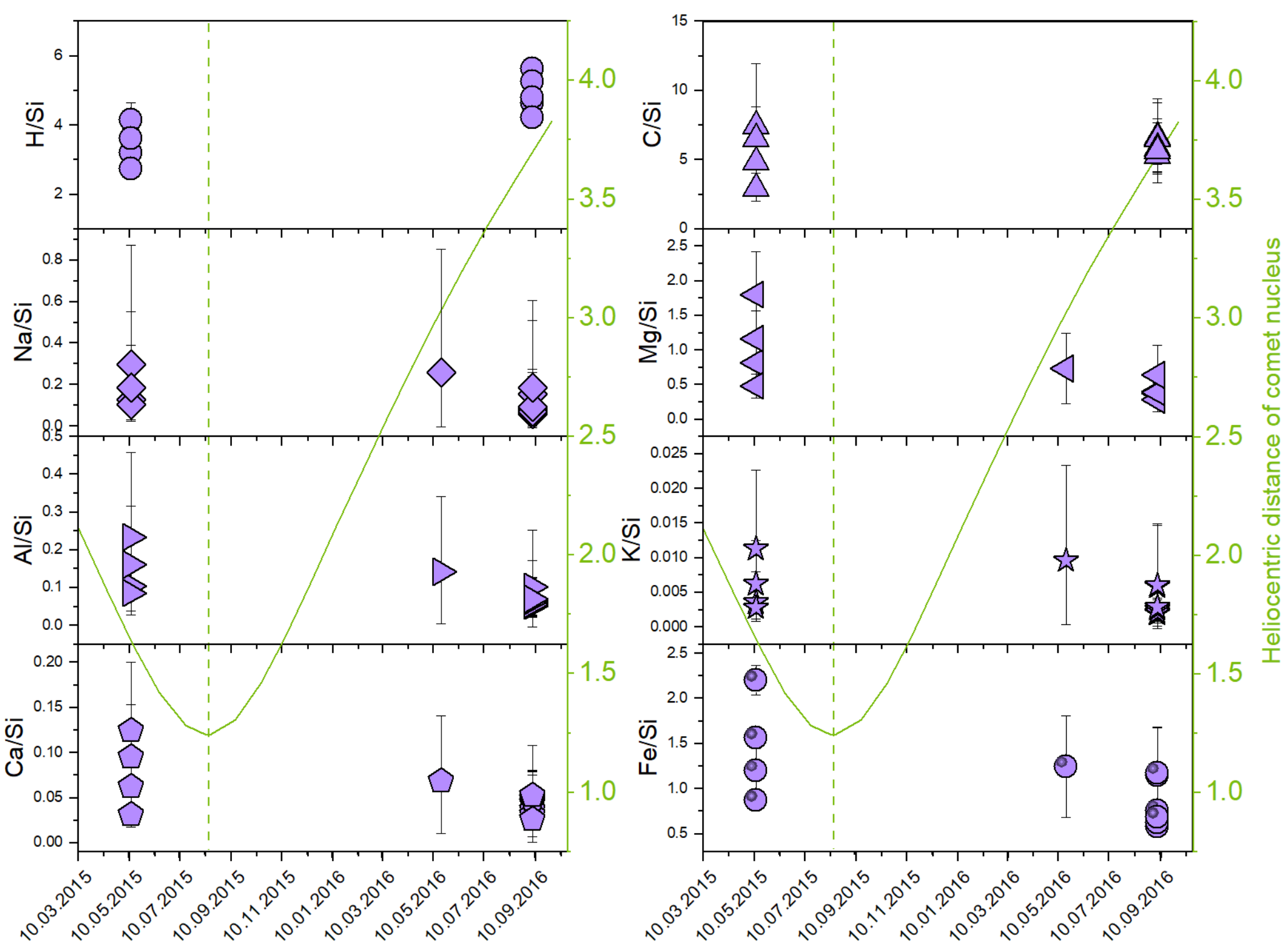

**Figure D 2.** Dust subsurface elemental ratios as a function of the heliocentric distance of the comet nucleus at the time of dust collection. In some cases, the error bars are not visible, as they are smaller than the symbol's size.

**Table 2:** List of the particles analyses used is this paper and ions counts for elements. (Complete table)

| N° | Target | Particle (First Name) | Particle (Last Name) | Starting date of the collection period | Ending date of the collection period) | Date of the Analysis | Number of considered spectra | H+ counts | ±σ | C+ counts | ±σ | Na+ counts | ±σ | Mg+ counts | ±σ | Al+ counts | ±σ | Si+ counts | ±σ | K+ counts | ±σ | Ca+ counts | ±σ | Fe+ counts | ±σ |
|---|---|---|---|---|---|---|---|---|---|---|---|---|---|---|---|---|---|---|---|---|---|---|---|---|---|
| | | | | | | | | ***No-sputtering analyses*** | | | | | | | | | | | | | | | | | |
| 1 | 1D0 | Sigrid(1) | Vesijako | 14/11/2014 | 21/11/2014 | 21/04/2016 | 50 | 8,06E+02 | 7,29E+00 | 1,19E+02 | 2,22E+00 | 6,95E+02 | 4,02E+00 | 1,13E+02 | 1,56E+00 | 2,23E+01 | 6,86E-01 | 1,38E+03 | 1,46E+01 | 5,48E+01 | 1,05E+00 | 2,06E+01 | 6,44E-01 | 1,12E+02 | 1,86E+00 |
| 2 | 2D0 | Estelle-Alizée(1) | Nilakka | 25/09/2014 | 03/10/2014 | 27/04/2016 | 56 | 5,19E+02 | 9,75E+00 | 6,68E+01 | 2,20E+00 | 2,31E+02 | 2,52E+00 | 3,70E+01 | 8,64E-01 | 2,21E+01 | 7,83E-01 | 9,40E+02 | 2,68E+01 | 2,39E+01 | 6,83E-01 | 6,86E+00 | 3,62E-01 | 4,17E+01 | 1,42E+00 |
| 3 | | Estelle-Alizée(2) | Nilakka | 25/09/2014 | 03/10/2014 | 28/04/2016 | 4 | 9,89E+01 | 5,05E+00 | 9,34E+00 | 1,56E+00 | 3,76E+01 | 3,08E+00 | 1,11E+01 | 1,68E+00 | -- | -- | 7,31E+01 | 4,50E+00 | -- | -- | -- | -- | 8,33E+00 | 1,47E+00 |
| 4 | 3D0 | Marius-Elias | Saimaa | 25/10/2014 | 31/10/2014 | 06/05/2016 | 6 | 5,40E+02 | 1,73E+01 | 1,14E+02 | 6,18E+00 | 2,44E+02 | 6,79E+00 | 1,95E+01 | 1,97E+00 | 1,90E+01 | 2,17E+00 | 6,81E+02 | 3,25E+01 | 5,17E+01 | 3,06E+00 | 1,35E+01 | 1,57E+00 | 5,09E+01 | 3,83E+00 |
| 5 | | Cleo | Ala-Kitka | 16/12/2014 | 20/12/2014 | 18/06/2016 | 1 | 7,00E+02 | 2,70E+01 | 1,05E+02 | 8,92E+00 | 9,07E+01 | 7,60E+00 | -- | -- | 4,07E+01 | 5,02E+00 | 1,33E+03 | 4,82E+01 | -- | -- | -- | -- | 7,58E+00 | 4,43E-01 |
| 6 | | Daniel | Kolima.3 | 24/01/2015 | 25/01/2015 | 17/06/2016 | 1 | 4,00E+02 | 2,17E+01 | 7,10E+01 | 8,31E+00 | 1,15E+02 | 8,64E+00 | 2,10E+01 | 3,44E+00 | -- | -- | 6,75E+02 | 3,55E+01 | 5,42E+01 | 5,46E+00 | -- | -- | 2,63E+01 | 5,35E+00 |
| 7 | | Erkka | Enonvesi | 20/12/2014 | 27/12/2014 | 18/06/2016 | 3 | 8,25E+02 | 2,76E+01 | 1,26E+02 | 7,76E+00 | 3,24E+02 | 8,40E+00 | 4,23E+01 | 2,80E+00 | 3,02E+01 | 2,47E+00 | 1,24E+03 | 5,71E+01 | 7,46E+01 | 3,66E+00 | 1,46E+01 | 1,68E+00 | 4,75E+01 | 4,31E+00 |
| 8 | | Fanny | Kolima.3 | 24/01/2015 | 25/01/2015 | 18/06/2016 | 1 | 4,45E+02 | 2,76E+01 | 7,85E+01 | 9,14E+00 | 1,09E+02 | 8,39E+00 | 1,49E+01 | 2,91E+00 | 5,26E+01 | 6,07E+00 | 1,04E+03 | 5,49E+01 | 6,40E+01 | 6,11E+00 | 1,47E+01 | 3,36E+00 | 2,46E+01 | 5,27E+00 |
| 9 | | Françoise | Ala-Kitka | 16/12/2014 | 20/12/2014 | 18/06/2016 | 1 | 5,98E+02 | 4,20E+01 | 1,24E+02 | 1,32E+01 | 5,84E+02 | 1,86E+01 | 3,05E+01 | 4,18E+00 | 6,66E+01 | 8,58E+00 | 9,83E+02 | 8,62E+01 | 1,18E+02 | 8,02E+00 | 2,42E+01 | 3,78E+00 | 5,06E+01 | 7,67E+00 |
| 10 | | Gervas | Ala-Kitka | 16/12/2014 | 20/12/2014 | 18/06/2016 | 2 | 6,74E+02 | 3,05E+01 | 1,11E+02 | 9,09E+00 | 2,80E+02 | 9,57E+00 | 2,61E+01 | 2,75E+00 | 2,04E+01 | 3,10E+00 | 1,39E+03 | 6,40E+01 | 7,54E+01 | 4,47E+00 | 1,24E+01 | 1,88E+00 | 3,18E+01 | 4,94E+00 |
| 11 | | Glenn | Lummene.4 | 28/01/2015 | 29/01/2015 | 17/06/2016 | 1 | 5,55E+02 | 2,46E+01 | 7,49E+01 | 8,96E+00 | 1,28E+02 | 8,96E+00 | 1,92E+01 | 3,34E+00 | -- | -- | 1,09E+03 | 4,00E+01 | 5,79E+01 | 6,02E+00 | 1,15E+01 | 3,00E+00 | 2,02E+01 | 5,13E+00 |
| 12 | | Jan | Kolima.3 | 24/01/2015 | 25/01/2015 | 18/06/2016 | 1 | 8,12E+02 | 3,77E+01 | 1,58E+02 | 1,21E+01 | 1,89E+02 | 1,08E+01 | 1,18E+01 | 2,64E+00 | 4,48E+01 | 6,23E+00 | 1,45E+03 | 7,53E+01 | 5,09E+01 | 5,31E+00 | 8,63E+00 | 2,25E+00 | 3,13E+01 | 5,76E+00 |
| 13 | 1CF | Joar | Lummene.4 | 28/01/2015 | 29/01/2015 | 18/06/2016 | 1 | 1,02E+03 | 4,41E+01 | 1,78E+02 | 1,39E+01 | 2,97E+02 | 1,36E+01 | 3,43E+01 | 4,41E+00 | 6,80E+01 | 6,51E+00 | 1,48E+03 | 8,50E+01 | 1,01E+02 | 8,06E+00 | 1,75E+01 | 3,40E+00 | 5,70E+01 | 7,70E+00 |
| 14 | | Katharina | Kolima.3 | 24/01/2015 | 25/01/2015 | 18/06/2016 | 2 | 5,29E+02 | 2,40E+01 | 8,57E+01 | 6,86E+00 | 1,03E+02 | 5,82E+00 | 1,43E+01 | 2,00E+00 | 2,29E+01 | 2,76E+00 | 1,00E+03 | 5,06E+01 | 3,44E+01 | 3,04E+00 | 8,07E+00 | 1,68E+00 | 1,92E+01 | 3,22E+00 |
| 15 | | Kenzi | Ala-Kitka | 16/12/2014 | 20/12/2014 | 17/06/2016 | 1 | 5,22E+02 | 2,26E+01 | 9,24E+01 | 9,19E+00 | 1,76E+02 | 1,02E+01 | 2,11E+01 | 3,59E+00 | -- | -- | 9,58E+02 | 3,44E+01 | 7,46E+01 | 6,38E+00 | 1,91E+01 | 3,81E+00 | 2,83E+01 | 5,73E+00 |
| 16 | | Khady | Ala-Kitka | 16/12/2014 | 20/12/2014 | 18/06/2016 | 1 | 1,35E+03 | 4,22E+01 | 2,14E+02 | 1,32E+01 | 4,40E+02 | 1,58E+01 | 7,95E+01 | 6,42E+00 | 3,43E+01 | 4,54E+00 | 1,45E+03 | 7,23E+01 | 6,43E+01 | 5,98E+00 | 2,13E+01 | 3,82E+00 | 9,53E+01 | 8,40E+00 |
| 17 | | Klaus | Inari | 02/01/2015 | 09/01/2015 | 18/06/2016 | 3 | 3,59E+02 | 1,19E+01 | 6,68E+01 | 4,44E+00 | 2,63E+01 | 2,66E+00 | 4,48E+00 | 9,96E-01 | 1,51E+01 | 2,21E+00 | 6,28E+02 | 2,01E+01 | 3,15E+01 | 2,37E+00 | 5,33E+00 | 1,01E+00 | -- | -- |
| 18 | | Lenz | Lummene.1 | 25/01/2015 | 26/01/2015 | 18/06/2016 | 2 | 1,27E+03 | 3,22E+01 | 1,88E+02 | 9,86E+00 | 3,40E+02 | 1,02E+01 | 3,51E+01 | 3,14E+00 | 5,36E+01 | 4,41E+00 | 1,97E+03 | 6,06E+01 | 8,86E+01 | 4,94E+00 | 2,23E+01 | 2,55E+00 | 6,50E+01 | 5,49E+00 |
| 19 | | Modou | Ala-Kitka | 16/12/2014 | 20/12/2014 | 18/06/2016 | 2 | 4,74E+02 | 1,74E+01 | 8,45E+01 | 5,96E+00 | 1,32E+02 | 6,20E+00 | 1,27E+01 | 1,89E+00 | 2,59E+01 | 3,04E+00 | 7,25E+02 | 3,07E+01 | 4,01E+01 | 3,31E+00 | 7,91E+00 | 1,50E+00 | 2,70E+01 | 3,28E+00 |
| 20 | | Silivia | Kolima.3 | 24/01/2015 | 25/01/2015 | 17/06/2016 | 2 | 5,17E+02 | 2,65E+01 | 8,13E+01 | 7,71E+00 | 8,05E+01 | 5,67E+00 | 1,14E+01 | 1,89E+00 | 2,01E+01 | 2,59E+00 | 1,24E+03 | 5,69E+01 | 7,05E+01 | 4,38E+00 | 7,49E+00 | 1,94E+00 | 2,32E+01 | 4,09E+00 |
| 21 | | Stavro | Kolima.3 | 24/01/2015 | 25/01/2015 | 17/06/2016 | 1 | 3,78E+02 | 2,74E+01 | 6,85E+01 | 8,96E+00 | 1,01E+02 | 8,15E+00 | 9,70E+00 | 2,51E+00 | 2,08E+01 | 3,50E+00 | 9,24E+02 | 5,59E+01 | 7,07E+01 | 6,29E+00 | 1,08E+01 | 2,84E+00 | 1,58E+01 | 4,78E+00 |
| 22 | | Alcide | Kolima.3 | 24/01/2015 | 25/01/2015 | 22/06/2016 | 2 | 4,39E+02 | 1,97E+01 | 6,94E+01 | 6,17E+00 | 1,97E+02 | 7,69E+00 | 2,48E+01 | 2,59E+00 | 1,83E+01 | 2,27E+00 | 9,13E+02 | 4,00E+01 | 3,21E+01 | 4,33E+00 | 6,66E+00 | 1,38E+00 | 3,05E+01 | 3,79E+00 |
| 23 | | Arlette | Kolima.3 | 24/01/2015 | 25/01/2015 | 22/06/2016 | 2 | 2,53E+02 | 1,29E+01 | 4,35E+01 | 4,38E+00 | 3,89E+01 | 3,50E+00 | 3,98E+00 | 1,08E+00 | 9,05E+00 | 1,89E+00 | 4,73E+02 | 2,43E+01 | 1,80E+01 | 2,20E+00 | 4,43E+00 | 1,28E+00 | 6,30E+00 | 1,96E+00 |
| 24 | | Birgit | Kolima.3 | 24/01/2015 | 25/01/2015 | 22/06/2016 | 2 | 9,04E+01 | 8,21E+00 | 1,48E+01 | 2,70E+00 | 4,78E+01 | 3,73E+00 | 4,11E+00 | 1,06E+00 | 4,74E+00 | 1,22E+00 | 1,61E+02 | 1,58E+01 | 1,10E+01 | 1,81E+00 | -- | -- | 9,15E+00 | 1,92E+00 |
| 25 | | Falco | Kolima.3 | 24/01/2015 | 25/01/2015 | 23/06/2016 | 2 | 5,37E+02 | 1,86E+01 | 8,29E+01 | 6,47E+00 | 2,01E+02 | 7,81E+00 | 2,81E+01 | 2,79E+00 | 2,11E+01 | 2,87E+00 | 7,50E+02 | 3,18E+01 | 5,02E+01 | 3,68E+00 | 4,30E+00 | 1,15E+00 | 1,83E+01 | 3,35E+00 |
| 26 | | Fred | Kolima.3 | 24/01/2015 | 25/01/2015 | 26/06/2016 | 3 | 3,65E+02 | 2,05E+00 | 5,11E+01 | 7,56E-01 | 2,64E+02 | 1,50E+00 | 4,07E+01 | 5,83E-01 | 9,13E+00 | 3,01E-01 | 5,96E+02 | 2,80E+00 | 1,89E+01 | 3,94E-01 | 5,62E+00 | 2,13E-01 | 3,76E+01 | 6,12E-01 |
| 27 | | Hartmut | Enonvesi | 20/12/2014 | 27/12/2014 | 22/06/2016 | 1 | 5,97E+02 | 2,83E+01 | 8,09E+01 | 9,18E+00 | 2,48E+02 | 1,21E+01 | 3,18E+01 | 4,20E+00 | 1,60E+01 | 3,23E+00 | 1,01E+03 | 5,14E+01 | 6,20E+01 | 5,71E+00 | 1,86E+01 | 3,70E+00 | 5,43E+01 | 6,59E+00 |
| 28 | | Hector | Kolima.3 | 24/01/2015 | 25/01/2015 | 22/06/2016 | 1 | 2,78E+02 | 1,90E+01 | 4,37E+01 | 6,20E+00 | 1,00E+02 | 7,57E+00 | 1,42E+01 | 2,72E+00 | 1,67E+01 | 3,42E+00 | 3,68E+02 | 3,41E+01 | 1,97E+01 | 3,77E+00 | -- | -- | 3,12E+01 | 4,68E+00 |
| 29 | | Irene | Ala-Kitka | 14/02/2015 | 14/02/2015 | 23/06/2016 | 1 | 6,73E+02 | 3,09E+01 | 1,10E+02 | 1,09E+01 | 1,98E+02 | 1,12E+01 | 2,59E+01 | 3,87E+00 | 1,62E+01 | 3,40E+00 | 1,56E+03 | 5,81E+01 | 6,47E+01 | 6,06E+00 | 1,36E+01 | 2,98E+00 | 3,36E+01 | 6,27E+00 |
| 30 | | Isko | Kolima.3 | 24/01/2015 | 25/01/2015 | 23/06/2016 | 1 | 1,03E+03 | 3,37E+01 | 1,73E+02 | 1,27E+01 | 2,70E+02 | 1,30E+01 | 3,04E+01 | 4,36E+00 | 2,55E+01 | 4,26E+00 | 1,54E+03 | 5,40E+01 | 9,80E+01 | 7,68E+00 | 1,51E+01 | 3,20E+00 | 8,15E+01 | 8,33E+00 |
| 31 | 2CF | Jean-Baptiste | Kolima.3 | 24/01/2015 | 25/01/2015 | 27/06/2016 | 65 | 8,97E+02 | 4,45E+00 | 1,39E+02 | 1,43E+00 | 3,76E+02 | 1,86E+00 | 4,92E+01 | 6,51E-01 | 1,74E+01 | 6,83E-01 | 1,54E+03 | 7,78E+00 | 3,67E+01 | 5,40E-01 | 1,18E+01 | 3,03E-01 | 6,64E+01 | 9,04E-01 |
| 32 | | Jessica | Lummene.2 | 26/01/2015 | 27/01/2015 | 06/04/2016 | 22 | 1,01E+03 | 1,13E+01 | 1,67E+02 | 3,55E+00 | 6,50E+02 | 5,80E+00 | 7,93E+01 | 1,94E+00 | 2,08E+01 | 1,11E+00 | 1,10E+03 | 2,16E+01 | 3,95E+01 | 1,38E+00 | 1,36E+01 | 7,91E-01 | 1,04E+02 | 2,51E+00 |
| 33 | | Kirsi | Enonvesi | 20/12/2014 | 27/12/2014 | 23/06/2016 | 2 | 1,12E+03 | 3,28E+01 | 1,87E+02 | 1,02E+01 | 3,14E+02 | 9,99E+00 | 3,15E+01 | 3,00E+00 | 2,20E+01 | 2,50E+00 | 1,74E+03 | 6,53E+01 | 6,29E+01 | 4,13E+00 | 1,28E+01 | 1,88E+00 | 7,46E+01 | 5,86E+00 |
| 34 | | Laurent | Enonvesi | 20/12/2014 | 27/12/2014 | 22/06/2016 | 3 | 6,17E+02 | 1,57E+01 | 1,01E+02 | 5,16E+00 | 2,03E+02 | 6,16E+00 | 2,69E+01 | 2,17E+00 | 1,41E+01 | 2,46E+00 | 9,18E+02 | 2,84E+01 | 2,62E+01 | 2,25E+00 | 6,24E+00 | 1,17E+00 | 4,10E+01 | 3,07E+00 |
| 35 | | Lumi | Kolima.3 | 24/01/2015 | 25/01/2015 | 23/06/2016 | 3 | 1,29E+03 | 3,36E+01 | 2,25E+02 | 1,02E+01 | 2,74E+02 | 8,19E+00 | 2,91E+01 | 2,50E+00 | 3,72E+01 | 2,67E+00 | 2,12E+03 | 6,99E+01 | 1,15E+02 | 4,48E+00 | 1,59E+01 | 1,67E+00 | 7,01E+01 | 5,37E+00 |
| 36 | | Regine | Kolima.3 | 24/01/2015 | 25/01/2015 | 23/06/2016 | 2 | 6,62E+02 | 1,97E+01 | 1,22E+02 | 7,17E+00 | 4,45E+02 | 1,11E+01 | 5,80E+01 | 3,91E+00 | 2,64E+01 | 3,29E+00 | 9,16E+02 | 3,31E+01 | 5,32E+01 | 3,83E+00 | 9,60E+00 | 1,68E+00 | 6,56E+01 | 4,83E+00 |
| 37 | | Roy | Kolima.3 | 24/01/2015 | 25/01/2015 | 04/03/2016 | 6 | 7,65E+02 | 2,54E+01 | 1,48E+02 | 7,70E+00 | 1,58E+02 | 6,06E+00 | 1,67E+01 | 1,81E+00 | 1,22E+01 | 1,83E+00 | 1,08E+03 | 5,33E+01 | 4,89E+01 | 2,97E+00 | 1,08E+01 | 1,39E+00 | 3,12E+01 | 3,54E+00 |
| 38 | | Seffi | Kolima.3 | 24/01/2015 | 25/01/2015 | 04/03/2016 | 4 | 6,68E+02 | 2,16E+01 | 1,19E+02 | 7,70E+00 | 1,76E+02 | 7,53E+00 | 1,47E+01 | 2,10E+00 | 1,14E+01 | 2,06E+00 | 1,04E+03 | 3,81E+01 | 5,71E+01 | 3,95E+00 | 1,72E+01 | 2,15E+00 | 1,91E+01 | 3,66E+00 |
| 39 | | Wibke | Kolima.3 | 24/01/2015 | 25/01/2015 | 23/06/2016 | 2 | 1,33E+03 | 3,47E+01 | 2,29E+02 | 1,15E+01 | 3,96E+02 | 1,14E+01 | 5,53E+01 | 4,02E+00 | 2,48E+01 | 3,00E+00 | 2,06E+03 | 6,54E+01 | 1,14E+02 | 5,59E+00 | 2,53E+01 | 2,63E+00 | 9,99E+01 | 6,91E+00 |
| 40 | | Zhong Yi | Enonvesi | 20/12/2014 | 27/12/2014 | 22/06/2016 | 3 | 4,40E+02 | 1,41E+01 | 6,77E+01 | 4,42E+00 | 1,85E+02 | 5,89E+00 | 2,00E+01 | 1,87E+00 | 1,30E+01 | 1,92E+00 | 7,83E+02 | 2,70E+01 | 2,39E+01 | 2,14E+00 | 8,28E+00 | 1,30E+00 | 5,41E+01 | 3,42E+00 |
| 41 | 3CF | Baudry | Kolima.3 | 24/01/2015 | 25/01/2015 | 05/02/2016 | 4 | 5,77E+02 | 1,23E+01 | 9,46E+01 | 4,92E+00 | 1,29E+01 | 1,89E+00 | -- | -- | -- | -- | 9,45E+02 | 1,59E+01 | 2,56E+01 | 2,73E+00 | 3,59E+00 | 1,17E+00 | 1,54E+01 | 2,14E+00 |
| 42 | | Claudia | Toisvesi.2 | 11/05/2015 | 12/05/2015 | 23/11/2015 | 8 | 1,92E+03 | 3,20E+01 | 3,55E+02 | 1,08E+01 | 1,32E+03 | 1,38E+01 | 1,59E+02 | 4,71E+00 | 5,71E+01 | 2,68E+00 | 2,64E+03 | 5,06E+01 | 1,46E+02 | 4,32E+00 | 3,40E+01 | 2,08E+00 | 2,08E+02 | 6,15E+00 |
| 43 | | David | Toisvesi.2 | 11/05/2015 | 12/05/2015 | 13/06/2015 | 24 | 1,51E+02 | 5,05E+00 | 2,26E+01 | 1,48E+00 | 6,21E+01 | 1,69E+00 | 9,65E+00 | 6,53E-01 | 1,12E+01 | 6,83E-01 | 1,70E+02 | 8,18E+00 | 1,17E+01 | 7,39E-01 | 1,55E+00 | 2,70E-01 | 1,08E+01 | 1,15E-01 |
| 44 | 2D1 | Jakub(2) | Toisvesi.2 | 11/05/2015 | 12/05/2015 | 13/06/2015 | 10 | 1,48E+02 | 4,80E+00 | 2,09E+01 | 1,66E+00 | 2,60E+01 | 1,65E+00 | 3,82E+00 | 6,42E-01 | 3,98E+00 | 6,38E-01 | 1,60E+02 | 5,98E+00 | 4,15E+00 | 7,48E-01 | 1,65E+00 | 4,63E-01 | 7,33E+00 | 9,75E-01 |
| 45 | | Kenneth | Toisvesi.2 | 11/05/2015 | 12/05/2015 | 18/06/2015 | 12 | 8,41E+02 | 1,22E+01 | 1,51E+02 | 4,45E+00 | 2,82E+02 | 5,03E+00 | 3,47E+01 | 1,77E+00 | 1,19E+01 | 1,24E+00 | 8,57E+02 | 1,64E+01 | 2,86E+01 | 1,61E+00 | 9,95E+00 | 9,47E-01 | 5,09E+01 | 2,32E+00 |
| 46 | | Roberto | Toisvesi.3 | 11/05/2015 | 12/05/2015 | 18/06/2015 | 4 | 2,97E+01 | 3,65E+00 | 3,74E+00 | 1,20E+00 | 1,41E+01 | 1,92E+00 | 3,45E+00 | 9,45E-01 | -- | -- | 3,98E+01 | 5,11E+00 | -- | -- | -- | -- | -- | -- |
| 47 | | Blat | Orivesi.4 | 31/07/2015 | 01/08/2015 | 16/10/2015 | 8 | 1,42E+03 | 2,87E+01 | 2,33E+02 | 8,40E+00 | 3,70E+02 | 9,70E+00 | 5,11E+01 | 2,77E+00 | 9,33E+00 | 1,20E+00 | 9,16E+02 | 3,55E+01 | 2,84E+01 | 1,96E+00 | 7,66E+00 | 1,02E+00 | 5,81E+01 | 3,41E+00 |
| 48 | | Bonin | Orivesi.4 | 31/07/2015 | 01/08/2015 | 21/08/2015 | 4 | 1,77E+03 | 4,53E+01 | 2,48E+02 | 1,35E+01 | 1,22E+03 | 2,22E+01 | 1,48E+02 | 6,55E+00 | 5,99E+01 | 5,07E+00 | 1,26E+03 | 5,69E+01 | 1,15E+02 | 5,44E+00 | 2,94E+01 | 2,74E+00 | 1,88E+02 | 8,12E+00 |
| 49 | | Bregan(1) | Orivesi.4 | 31/07/2015 | 01/08/2015 | 21/08/2015 | 4 | 1,23E+03 | 3,27E+01 | 2,29E+02 | 1,18E+01 | 7,31E+02 | 1,73E+01 | 9,46E+01 | 5,36E+00 | 7,72E+01 | 5,52E+00 | 8,74E+02 | 3,86E+01 | 1,20E+02 | 5,76E+00 | 2,35E+01 | 2,53E+00 | 1,28E+02 | 6,83E+00 |
| 50 | 1CD | Bregan(2) | Orivesi.4 | 31/07/2015 | 01/08/2015 | 10/09/2015 | 29 | 1,37E+03 | 1,19E+01 | 2,12E+02 | 4,13E+00 | 6,25E+02 | 5,89E+00 | 8,78E+01 | 1,93E+00 | 4,01E+01 | 1,91E+00 | 2,17E+03 | 1,53E+01 | 9,18E+01 | 1,82E+00 | 1,95E+01 | 8,28E-01 | 1,10E+02 | 2,37E+00 |
| 51 | | Bregan(3) | Orivesi.4 | 31/07/2015 | 01/08/2015 | 19/02/2016 | 27 | 1,36E+03 | 1,21E+01 | 1,89E+02 | 4,00E+00 | 3,26E+02 | 4,77E+00 | 4,39E+01 | 1,47E+00 | 1,77E+01 | 1,77E+00 | 3,01E+03 | 1,73E+01 | 6,06E+01 | 1,52E+00 | 1,31E+01 | 7,17E-01 | 5,94E+01 | 1,91E+00 |
| 52 | | Dedan | Orivesi.4 | 31/07/2015 | 01/08/2015 | 16/01/2016 | 3 | 9,51E+02 | 5,94E+01 | 1,55E+02 | 1,59E+01 | 3,01E+02 | 1,82E+01 | 3,65E+01 | 4,26E+00 | 1,21E+01 | 2,45E+00 | 6,53E+02 | 8,20E+01 | 4,41E+01 | 4,04E+00 | -- | -- | 3,39E+01 | 5,61E+00 |
| 53 | | Devoll | Orivesi.4 | 31/07/2015 | 01/08/2015 | 16/10/2015 | 4 | 2,89E+02 | 1,48E+01 | 4,04E+01 | 4,64E+00 | 8,33E+01 | 5,86E+00 | 1,68E+01 | 2,18E+00 | 4,95E+00 | 1,26E+00 | 5,27E+02 | 2,11E+01 | 8,43E+00 | 1,67E+00 | -- | -- | 1,06E+01 | 2,07E+00 |
| 54 | | Umeka | Jonkeri.2 | 03/07/2015 | 04/07/2015 | 10/09/2015 | 6 | 1,50E+03 | 2,65E+01 | 2,46E+02 | 8,90E+00 | 3,96E+02 | 1,05E+01 | 4,87E+01 | 3,14E+00 | 1,17E+01 | 1,59E+00 | 1,92E+03 | 3,41E+01 | 5,07E+01 | 2,98E+00 | 1,07E+01 | 1,40E+00 | 6,31E+01 | 4,02E+00 |
| 55 | | Juliette | Hankavesi.1 | 23/10/2015 | 29/10/2015 | 18/11/2015 | 46 | 8,30E+02 | 7,13E+00 | 1,39E+02 | 2,33E+00 | 5,99E+02 | 3,88E+00 | 5,53E+01 | 1,13E+00 | 1,25E+01 | 5,87E-01 | 1,13E+03 | 1,21E+01 | 3,17E+01 | 8,40E-01 | 1,09E+01 | 4,92E-01 | 1,03E+02 | 1,62E+00 |
| 56 | | Fadil | Kolima.2 | 16/11/2015 | 18/11/2015 | 26/11/2015 | 10 | 4,10E+02 | 1,16E+01 | 6,49E+01 | 3,58E+00 | 1,84E+02 | 4,70E+00 | 2,18E+01 | 1,52E+00 | 1,25E+01 | 1,17E+00 | 8,01E+02 | 2,28E+01 | 2,63E+01 | 1,70E+00 | 2,25E+00 | 5,61E-01 | 3,69E+01 | 2,12E+00 |
| 57 | | Günter(1) | Jerisjarvi.1 | 29/02/2016 | 01/03/2016 | 15/04/2016 | 100 | 4,44E+02 | 2,86E+00 | 6,14E+01 | 9,35E-01 | 1,63E+02 | 1,35E+00 | 2,65E+01 | 5,25E-01 | 7,90E+00 | 2,95E-01 | 7,22E+02 | 4,59E+00 | 1,08E+01 | 3,36E-01 | 4,19E+00 | 2,08E-01 | 3,63E+01 | 6,31E-01 |
| 58 | 1D2 | Günter(2) | Jerisjarvi.1 | 29/02/2016 | 01/03/2016 | 08/05/2016 | 101 | 5,06E+02 | 2,76E+00 | 6,39E+01 | 9,12E-01 | 1,16E+02 | 1,12E+00 | 1,74E+01 | 4,24E-01 | 4,88E+00 | 2,58E-01 | 8,55E+02 | 4,07E+00 | 1,00E+01 | 3,22E-01 | 3,25E+00 | 1,88E-01 | 2,43E+01 | 5,23E-01 |
| 59 | | Enton | Hankavesi.1 | 23/10/2015 | 29/10/2015 | 20/11/2015 | 2 | 8,88E+02 | 2,91E+01 | 1,22E+02 | 1,05E+01 | 6,54E+02 | 1,89E+01 | 5,22E+01 | 5,27E+00 | 1,92E+01 | 3,95E+00 | 1,44E+03 | 4,48E+01 | 5,54E+01 | 5,63E+00 | 1,10E+01 | 2,59E+00 | 8,90E+01 | 7,30E+00 |
| 60 | | Erzan | Kolima.2 | 16/11/2015 | 18/11/2015 | 27/11/2015 | 4 | 6,52E+02 | 1,71E+01 | 1,03E+02 | 6,27E+00 | 2,13E+02 | 7,88E+00 | 4,72E+01 | 3,50E+00 | 1,33E+01 | 2,08E+00 | 9,07E+02 | 2,57E+01 | 3,21E+01 | 3,00E+00 | 4,69E+00 | 1,22E+00 | 4,35E+01 | 3,76E+00 |
| 61 | | Sophie | Kolima.2 | 16/11/2015 | 18/11/2015 | 26/11/2015 | 9 | 1,33E+02 | 6,06E+00 | 1,56E+01 | 1,73E+00 | 1,74E+01 | 1,56E+00 | 3,27E+00 | 6,21E-01 | 4,33E+00 | 8,79E-01 | 3,46E+02 | 1,22E+01 | 3,09E+00 | 6,90E-01 | -- | -- | 3,86E+00 | 8,64E-01 |

| | | | | | | | | | | | | | | | | |
|---|---|---|---|---|---|---|---|---|---|---|---|---|---|---|---|---|
| 62 | 2D2 | Stefan | Ukonvesi.6 | 17/01/2016 | 18/01/2016 | 11/02/2016 | 6 | 1,35E+03 4,57E+01 | 2,31E+02 1,22E+01 | 1,71E+02 6,25E+00 | 2,18E+01 2,05E+00 | 1,97E+01 2,05E+00 | 2,12E+03 8,95E+01 | 5,03E+01 2,98E+00 | 7,51E+00 1,19E+00 | 5,50E+01 3,74E+00 |
| 63 | 3C3 | Herbert_Hernest | Ala-Kitka.5 | 12/05/2016 | 13/05/2016 | 09/06/2016 | 8 | 9,51E+02 3,76E+01 | 1,97E+02 1,16E+01 | 2,40E+02 5,92E+00 | 3,64E+01 2,39E+00 | 3,77E+01 2,49E+00 | 1,07E+03 6,39E+01 | 1,33E+02 4,15E+00 | 3,21E+01 2,06E+00 | 4,65E+01 3,45E+00 |
| Analyses number considered per ions | | | | | | | | 63 | 63 | 63 | 61 | 57 | 63 | 60 | 54 | 61 |
| ***Pre-sputtering*** | | | | | | | | | | | | | | | | |
| 1 | 1C3 | Charlotte | Vesijako.1 | 05/09/2016 | 11/09/2016 | 13/09/2016 | 2 | 1,35E+02 4,17E+00 | 3,13E+01 1,79E+00 | 2,23E+01 1,43E+00 | 3,11E+00 5,30E-01 | 1,71E+00 4,45E-01 | 1,96E+02 6,10E+00 | 1,71E+00 4,24E-01 | -- -- | 4,31E+00 6,99E-01 |
| 2 | | Cartsen | Vesijako.1 | 05/09/2016 | 11/09/2016 | 13/09/2016 | 2 | 9,04E+01 3,04E+00 | 2,50E+01 1,51E+00 | 1,11E+01 9,79E-01 | 8,96E-01 2,83E-01 | -- -- | 1,30E+02 4,00E+00 | 1,48E+00 3,83E-01 | -- -- | 2,18E+00 4,83E-01 |
| 3 | | Julia | Vesijako.1 | 05/09/2016 | 11/09/2016 | 13/09/2016 | 2 | 2,19E+02 5,21E+00 | 5,07E+01 2,27E+00 | 6,41E+01 2,34E+00 | 7,41E+00 7,91E-01 | 2,53E+00 4,73E-01 | 3,01E+02 7,33E+00 | 3,59E+00 5,98E-01 | 1,92E+00 4,71E-01 | 1,31E+01 1,19E+00 |
| 4 | | Lou | Enonvesi.7 | 19/05/2016 | 19/05/2016 | 13/09/2016 | 6 | 1,89E+02 2,71E+00 | 5,58E+01 1,35E+00 | 3,87E+01 1,06E+00 | 5,47E+00 3,96E-01 | 1,44E+00 2,11E-01 | 2,21E+02 3,43E+00 | 2,53E+00 2,75E-01 | 1,51E+00 2,29E-01 | 9,68E+00 5,75E-01 |
| 5 | 2D1 | Robert | Toisvesi.2 | 11/05/2015 | 12/05/2015 | 08/09/2016 | 1 | 1,12E+02 4,67E+00 | 2,67E+01 2,13E+00 | 9,08E+01 3,43E+00 | 9,31E+00 1,10E+00 | 2,11E+00 6,11E-01 | 8,85E+01 5,59E+00 | 3,35E+00 7,13E-01 | -- -- | 1,24E+01 1,36E+00 |
| 6 | | Stephan | Toisvesi.2 | 11/05/2015 | 12/05/2015 | 08/09/2016 | 1 | 1,63E+02 4,95E+00 | 4,26E+01 2,42E+00 | 3,71E+01 2,17E+00 | 6,30E+00 8,97E-01 | 2,10E+00 5,41E-01 | 1,76E+02 5,71E+00 | 2,41E+00 6,38E-01 | -- -- | 1,20E+01 1,27E+00 |
| 7 | | Filipe | Toisvesi.2 | 11/05/2015 | 12/05/2015 | 08/09/2016 | 1 | 5,07E+01 3,58E+00 | 1,24E+01 1,57E+00 | 3,80E+01 2,24E+00 | 3,98E+00 7,25E-01 | 1,92E+00 5,43E-01 | 2,26E+01 4,50E+00 | 2,92E+00 6,51E-01 | -- -- | 5,96E+00 9,87E-01 |
| 8 | | Andrea | Toisvesi.2 | 11/05/2015 | 12/05/2015 | 08/09/2016 | 1 | 5,27E+00 1,43E+00 | 2,15E+00 6,97E-01 | 1,87E+01 1,56E+00 | 1,60E+00 4,57E-01 | -- -- | -- -- | 2,11E+00 5,23E-01 | -- -- | 2,64E+00 6,72E-01 |
| Analyses number considered per ions | | | | | | | | 8 | 8 | 8 | 8 | 6 | 7 | 8 | 2 | 8 |
| ***Post-sputtering*** | | | | | | | | | | | | | | | | |
| 1 | 1C3 | Charlotte(A1) | Vesijako.1 | 05/09/2016 | 11/09/2016 | 16/09/2016 | 4 | 8,76E+01 2,46E+00 | 1,56E+01 9,74E-01 | 2,27E+02 3,26E+00 | 1,14E+02 2,22E+00 | 1,95E+01 9,12E-01 | 1,08E+02 2,81E+00 | 1,41E+01 7,78E-01 | 2,55E+01 1,13E+00 | 9,15E+01 2,03E+00 |
| 2 | | Charlotte(A2) | Vesijako.1 | 05/09/2016 | 11/09/2016 | 16/09/2016 | 2 | 3,86E+01 1,99E+00 | 6,29E+00 7,84E-01 | 1,15E+02 3,15E+00 | 7,02E+01 2,43E+00 | 9,36E+00 8,90E-01 | 4,76E+01 2,24E+00 | 7,63E+00 8,09E-01 | 1,74E+01 1,24E+00 | 4,89E+01 2,04E+00 |
| 3 | | Cartsen(A1) | Vesijako.1 | 05/09/2016 | 11/09/2016 | 16/09/2016 | 4 | 6,14E+01 1,75E+00 | 1,05E+01 7,00E-01 | 1,81E+02 2,79E+00 | 8,94E+01 1,94E+00 | 1,33E+01 7,50E-01 | 6,33E+01 1,80E+00 | 7,65E+00 5,76E-01 | 2,31E+01 1,00E+00 | 5,00E+01 1,47E+00 |
| 4 | | Cartsen(A2) | Vesijako.1 | 05/09/2016 | 11/09/2016 | 17/09/2016 | 2 | 3,56E+01 1,79E+00 | 6,30E+00 7,45E-01 | 1,18E+02 3,15E+00 | 5,75E+01 2,20E+00 | 8,19E+00 8,38E-01 | 3,88E+01 1,88E+00 | 4,82E+00 6,45E-01 | 1,30E+01 1,05E+00 | 3,28E+01 1,67E+00 |
| 5 | | Cartsen(A3) | Vesijako.1 | 05/09/2016 | 11/09/2016 | 19/09/2016 | 1 | -- -- | -- -- | 4,39E+01 2,99E+00 | 1,22E+01 1,47E+00 | 2,90E+00 7,02E-01 | 8,08E+00 2,49E+00 | 3,02E+00 7,22E-01 | 2,25E+00 6,83E-01 | 1,26E+01 1,50E+00 |
| 6 | | Julia(A1) | Vesijako.1 | 05/09/2016 | 11/09/2016 | 16/09/2016 | 4 | 7,35E+01 2,02E+00 | 1,36E+01 8,57E-01 | 3,17E+02 3,75E+00 | 1,45E+02 2,48E+00 | 2,19E+01 9,65E-01 | 9,74E+01 2,33E+00 | 1,75E+01 8,60E-01 | 1,93E+01 9,57E-01 | 8,99E+01 1,99E+00 |
| 7 | | Julia(A2) | Vesijako.1 | 05/09/2016 | 11/09/2016 | 19/09/2016 | 2 | 4,84E+00 1,13E+00 | -- -- | 2,96E+01 1,71E+00 | 1,09E+01 9,81E-01 | 1,09E+00 3,13E-01 | 4,51E+00 1,23E+00 | 1,74E+00 3,98E-01 | 1,79E+00 4,18E-01 | 7,19E+00 8,16E-01 |
| 8 | | Lou(A) | Enonvesi.7 | 19/05/2016 | 19/05/2016 | 19/09/2016 | 1 | 1,04E+01 3,13E+00 | -- -- | 1,05E+02 4,72E+00 | 3,18E+01 2,41E+00 | 5,72E+00 9,92E-01 | 1,15E+01 3,61E+00 | 7,03E+00 1,15E+00 | 5,97E+00 1,08E+00 | 1,93E+01 1,96E+00 |
| 9 | 2D1 | Robert(A) | Toisvesi.2 | 11/05/2015 | 12/05/2015 | 10/09/2016 | 1 | 2,04E+02 7,98E+00 | 3,21E+01 2,68E+00 | 4,62E+03 2,49E+01 | 3,03E+03 1,97E+01 | 3,63E+02 6,80E+00 | 4,45E+02 1,20E+01 | 3,21E+02 6,37E+00 | 4,21E+02 7,74E+00 | 1,32E+03 1,41E+01 |
| 10 | | Stephan(A) | Toisvesi.2 | 11/05/2015 | 12/05/2015 | 10/09/2016 | 1 | 1,83E+02 5,89E+00 | 3,39E+01 2,33E+00 | 1,86E+03 1,56E+01 | 1,25E+03 1,26E+01 | 1,59E+02 4,52E+00 | 2,83E+02 7,65E+00 | 1,12E+02 3,77E+00 | 2,04E+02 5,29E+00 | 5,98E+02 8,93E+00 |
| 11 | | Filipe(A) | Toisvesi.2 | 11/05/2015 | 12/05/2015 | 10/09/2016 | 1 | 1,49E+02 4,61E+00 | 3,58E+01 2,20E+00 | 8,00E+02 1,01E+01 | 4,02E+02 7,14E+00 | 6,66E+01 2,90E+00 | 2,21E+02 5,66E+00 | 3,92E+01 2,25E+00 | 5,22E+01 2,64E+00 | 2,61E+02 5,78E+00 |
| 12 | | Andrea(A) | Toisvesi.2 | 11/05/2015 | 12/05/2015 | 10/09/2016 | 1 | 1,45E+01 1,66E+00 | 3,24E+00 7,06E-01 | 7,80E+01 3,24E+00 | 5,38E+01 2,62E+00 | 6,41E+00 9,11E-01 | 1,75E+01 1,95E+00 | 3,95E+00 7,21E-01 | 8,27E+00 1,06E+00 | 2,83E+01 1,93E+00 |
| Analyses number considered per ions | | | | | | | | 11 | 9 | 12 | 12 | 12 | 12 | 12 | 12 | 12 |